\documentclass{amsart}
\usepackage{enumitem}
\usepackage[all]{xy}
\usepackage[margin=1.00in, marginparwidth=1.5cm, marginparsep=0.5cm]{geometry}

\usepackage[section]{placeins}

\usepackage{booktabs}

\usepackage{lipsum} 

\usepackage{verbatim}

\usepackage{mathtools}

\usepackage{bbold}

\usepackage{amssymb}
\usepackage{array}
\usepackage{xcolor}

\newcolumntype{C}[1]{>{\centering\arraybackslash}p{#1}}

\usepackage{mathrsfs}
\usepackage{physics}
\usepackage{amsthm}
\usepackage{amsmath,amsfonts,graphicx,mathdots,cite,wrapfig}
\usepackage[colorlinks,linkcolor=blue,citecolor=blue,pagebackref,hypertexnames=false, breaklinks]{hyperref}
\usepackage{cite}
\usepackage{mwe}
\usepackage{caption}

\usepackage{subfigure} 
\usepackage{marginnote}

\newtheorem{theorem}{Theorem}[section]
\newtheorem{lemma}[theorem]{Lemma}
\newtheorem{proposition}[theorem]{Proposition}
\newtheorem{corollary}[theorem]{Corollary}

\theoremstyle{definition}
\newtheorem{definition}[theorem]{Definition}

\theoremstyle{remark}
\newtheorem{remark}[theorem]{Remark}

\numberwithin{equation}{section}

\DeclareMathOperator{\integers}{{\mathbb{Z}}}
\DeclareMathOperator{\nn}{\mathbb{N}}

\DeclareMathOperator{\complex}{\mathbb{C}}

\DeclareMathOperator{\mult}{mult}

\DeclareMathOperator{\dom}{Dom}

\begin{document}


\title{Gaps Labeling Theorem for the Bubble-Diamond Self-similar Graphs}


\author[Melville]{Elizabeth Melville}
\address{Elizabeth Melville, Mathematics Department, Brigham Young University, Provo, UT 84602}
\email{melville@mathematics.byu.edu}
\author[Mograby]{Gamal Mograby}
\address{Mathematics Department, Tufts University, Medford, MA 02155}
\email{gamal.mograby@uconn.edu}
\author[Nagabandi]{Nikhil Nagabandi}
\address{Nikhil Nagabandi, Mathematics Department, University of North Carolina, Chapel Hill, NC 27599}
\email{nikhil24@email.unc.edu}
\author[Rogers]{Luke~G. Rogers}
\address{Luke~G. Rogers, Mathematics Department, University of Connecticut, Storrs, CT 06269}
\email{luke.rogers@uconn.edu}
%
%
\author[Teplyaev]{Alexander Teplyaev}
\address{Alexander Teplyaev, Mathematics \& Physics Department, University of Connecticut, Storrs, CT 06269}
\email{teplyaev@uconn.edu}
%
%


\date{\today}


\begin{abstract}
Motivated by the appearance of fractals in several areas of physics, especially in solid state physics and the physics of aperiodic order, and in other sciences, including the quantum information theory, we present a detailed spectral analysis for a new class of fractal-type diamond graphs, referred to as bubble-diamond graphs, and provide a gap-labeling theorem in the sense of Bellissard for the corresponding probabilistic graph Laplacians using the technique of spectral decimation. Labeling the gaps in the Cantor set by the normalized eigenvalue counting function, also known as the integrated density of states, we describe the gap labels as  orbits of a second dynamical system that reflects the branching parameter of the bubble construction and the decimation structure. The spectrum of the natural Laplacian on limit graphs is shown generically to be pure point  supported on a Cantor set, though   one particular graph has a mixture of pure point and singularly continuous components.
\end{abstract}

\maketitle

\tableofcontents

\section{Introduction}
Fractals appeared in many prominent  physics papers in the past half-century, see 
\cite{Englert1987,Reuter2005,Rammal1983,RammalToulouse1983,Rammal1984,BellissardRenormalizationGroup,BellissardBovierGhez1992,Alexander1983,Mandelbrot1981}
for some of the foundational work most relvant to our article. Recently some fractal structures were analyzed in relation to quantum information theory \cite{pqst1,pqst2}. Motivated by these, this paper presents a detailed spectral analysis for a new class of fractal-type diamond graphs, referred to as \textit{bubble-diamond} graphs.
We investigate the Bubble-diamond graphs as a family of self-similar graphs for which the interplay between graph topology and spectral gaps labeling is transparent and explicit. The class of fractal-type diamond graphs is interesting in part because it is large and diverse, including examples for which the  limit spaces have a wide range of Hausdorff and spectral dimensions. 
The structure of these graphs is such that they combine spectral properties of Dyson hierarchical models and transport properties of one dimensional chains \cite{Akkermans2009ComplexDim,HamblyKumagai2010,NekrashevychTeplyaev2008,AlonsoRuiz2018,Teplyaev2008HarmonicCoordinates,Brzoska2017Magnetic}. 
Gap-labeling theorems are significant in solid-state physics, spectral analysis and K-theory. Historically, the discovery that certain Schr{\"o}dinger-type operators have Cantor spectra \cite{Harper_1955, DinaburgSinai1975, Moser1981,  Hofstadter1976} was clarified in the work of Bellissard, who  observed that the spectral gaps in these Cantor sets could be labeled using the integrated density of states of the Schr{\"o}dinger operator  \cite{BellissardNTBook1992, BellissardBovierGhez1992,BellissardNoncommutative2003}. The gap values of the integrated density of states lie in a specific countable set of numbers which are rigid under small perturbations of the Schr{\"o}dinger  operator, and Bellissard determined that this stability has a topological nature.

The bubble-diamond graphs are defined in Section~\ref{section:BubbleDiamond}. There are several ways to define these graphs, involving either edge branching~\cite{AlonsoRuiz2018} or substitution of edges of a graph by copies of another graph~\cite{MalozemovTeplyaev2003}, both of which are related to the fractafold constructions in~\cite{BobFractafolds2003,BobTransfOfspectra2010, StrichartzTeplyaev2012}. 
Intuitively, we take a basic building block graph $G_1$  and inductively replace the edges of $G_1$ by copies of $G_\ell$ to obtain $G_{\ell+1}$, see Figures~\ref{fig:bubblediffBranch} and~\ref{fig:copiesLabel}. Identifying $G_\ell$ with one of the copies in $G_{\ell+1}$ we can let $G_\infty=\cup_\ell G_\ell$. The limit depends on the identifications, which we restrict in a manner that ensures $G_\infty$ is regular locally finite graph without boundary (see Theorem~\ref{thm:purepointSpectrum}).

We equip $G_\infty$ and each $G_\ell$ with a Hilbert space $L^2(G_{\infty})$ (resp. $L^2(G_{\ell})$) and a  probabilistic graph Laplacian $\Delta_{\infty,b}$ (resp. $\Delta_{\ell,b}$). In the finite graph case, we also consider the Dirichlet graph Laplacian $\Delta^D_{\ell,b}$. In Section~\ref{section:BubbleDiamond} we use a minor variation on the proof of  Theorem~5.8 in~\cite{MalozemovTeplyaev2003} to show that the spectrum $\sigma(\Delta_{\infty,b})$ is the closure of the set of all Dirichlet eigenvalues of $\Delta^D_{\ell,b}$, $\ell \geq 1$ because Dirichlet-Neumann eigenfunctions of $\Delta^D_{\ell,b}$ can be extended to  compactly supported eigenfunctions of $\Delta_{\infty,b}$ that form a complete set in $L^2(G_{\infty})$. Each eigenvalue in $\sigma(\Delta_{\infty,b})$ is of infinite multiplicity.  We note that Dirichlet-Neumann eigenfunctions occur for self-similar graphs and fractals with sufficient symmetry and dramatically simplify the spectral analysis~\cite{MalozemovTeplyaev1995,BarlowKigami1997, Sabot2000,KigamiBook2001,Teplyaev1998}.

In Section~\ref{sec-spectralDeci}, we apply a technique common in analysis on fractals called the \textit{spectral decimation method} to relate the spectra of the Laplacians $\Delta_{\ell,b}$ via iteration of a rational function $R_b$ and show $\Delta_{\infty,b}$ is the Julia set $\mathcal{J}(R_b)$. This method has a long history~\cite{Rammal1984,RammalToulouse1983,BajorinVibration3Ngasket2008,BajorinVibrationSpectra2008,KigamiBook2001,BellissardRenormalizationGroup,ShimaFukushima1992,ShimaPreSierpinski1991,BobBook2006,Cao2022}.
The spectral decimation function $R_b$ is computed to be a cubic polynomial with critical points in the basin of attraction of $\infty$, so by Theorem~13.1(2) of~\cite{Brolin1965} its Julia set is totally disconnected and Lebesgue measure zero.  $R_b$ coincides with the spectral decimation function for the \textit{$pq$-model} investigated in~\cite{ChenTeplyaev2016} (with $p=\frac{b}{b+1}$ or $p=\frac{1}{b+1}$), so the spectrum of $\Delta_{\infty,b}$ coincides as a set with that of the Laplacian in the \textit{$pq$-model}; nevertheless, their spectral types are different, with pure point spectrum possibly mixed with singularly continuous spectrum, see Theorem~\ref{thm:purepointSpectrum} and Theorem~\ref{prop:sc}. This is a rare spectral feature previously   observed only in the case of the Sierpinski gaskets, in which setting the proofs are   much more complicated~\cite{Teplyaev1998,quint}.

Section~\ref{sec-IDS} is concerned with the structure of the spectrum of  $\Delta_{\infty,b}$. We show the density of states measure $\nu_b$ is atomic with support the Julia set $\mathcal{J}(R_b)$, give an explicit formula for it~\eqref{eq:defnDOS}, and characterize its self-similar structure~\eqref{eq:DOSselfsimilar}. In Section~\ref{sec-GapLabeling} we exploit this latter self-similarity to see that the normalized eigenvalue counting function $N_b(x)=\nu_b((-\infty,x])$ also has a self-similar property.  This provides the main results of the paper: Theorem~\ref{thm:gaplabeling}, which explicitly identifies the values of $N_b(x)$ on the the gaps in the spectrum $\sigma(\Delta_{\infty,b})$, which  are Bellissard's gap labels in this setting, and Corollary~\ref{cor:dynamicalgaplabeling} which describes these gap labels as the orbits of a dynamical system and the range of $N_b$ as the associated Julia set.   One potentially useful observation is that the collection of rational numbers that occur as gap labels have denominators that reflect the number of self-similar copies of the fractal graph.   It should be noted that this method of gap labeling via orbits of a dynamical system is quite generally available in settings where there is spectral decimation, such as those in~\cite{BajorinVibration3Ngasket2008,BajorinVibrationSpectra2008},  but to the best of our knowledge this is the first time it has appeared in the literature.  Our results are closely related to recent research~\cite{BaluMogOkoTep2021spectralAMO,BaluMogOkoTepJacobi2022}. The aim of our paper is provide a useful model for further investigation.

Finally, in Section~\ref{sec:Sabotsection0} we note the implications for the situation where one renormalizes the graphs to converge to a compact fractal limit.  The details involved in taking this limit are mostly standard so are only sketched; the main result is that for the Laplacian on this compact limit one has gap sequences in the sense of~\cite{StrichartzGaps} and that the natural measurement of the sizes of these gap sequences can be computed from the gap labels in Section~\ref{sec-GapLabeling} and the Koenigs linearization of the inverse branch of $R_b$ at the fixed point $0$.

Our work is part of a long term study of  mathematical physics on fractals and self-similar graphs  \cite{dumitrescu2022dynamical,KassoStrichartz2005,KassoStrichartz2007,KassoSaloffCosteTeplyaev2008,Akkermans2009ComplexDim,Akkermans2012ThermoPhoton,Akkermans2012SpatialLogPeriodic,Akkermans2020AC_circuits,Akkermans2013quantumFieldsFractals,Dunne2012HeatKernels,
AlonsoRuiz2016Hanoi,HinzMeinert2020,KigamiLapidus1993}, in which novel features of physical systems on fractals can be associated with the unusual spectral and geometric properties of fractals compared to regular graphs and smooth manifolds.

\section{Bubble-diamond Self-similar Graphs}\label{section:BubbleDiamond}

We define a sequence of graphs which are related to the diamond fractals of\cite{AlonsoRuiz2018} and are a special case of the construction in~\cite{MalozemovTeplyaev2003}. They depend on a branching parameter $b$ which gives the number of branches that form the bubble.

\begin{figure}[htb]
        \centering
        \includegraphics[width=0.8\linewidth]{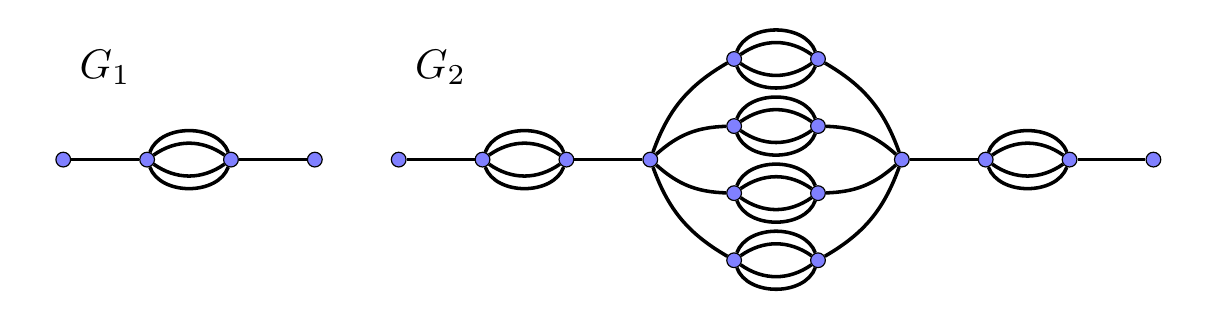}
    	  \caption{Level $G_1$ and $G_2$ in the case $b=4$.}
    	  \label{fig:bubblediffBranch}
\end{figure}

For any graph $G$ we denote the set of edges $E(G)$ and the set of vertices $V(G)$.
\begin{definition}\label{def:BubbleDiamondGraphs}
Fix $b \geq 2$, $b \in \nn$. The  \textit{Bubble diamond graphs (with branching parameter $b$)} are $\{G_{\ell}\}_{\ell \geq 0}$ constructed inductively.  $G_{0}$ has two vertices joined by an edge.  At level $\ell$ we construct $G_{\ell}$ by modifying each edge from  $G_{\ell-1}$ as follows: introduce two new vertices, each of which is joined to one of the two original vertices by an edge and which are joined to one another by $b$ distinct edges. Write $\deg_{\ell}(p)$ for the degree of a vertex $p$ and $\partial G_{\ell} =\{ p \in V(G_{\ell}): \deg_{\ell}(p) = 1 \}$ for the boundary vertices.
\end{definition}
The steps from $G_1$ to $G_2$ are shown for branching parameter $b=4$ in Figure~\ref{fig:bubblediffBranch}.  Note that we consider the branching parameter to be fixed throughout this work and suppress the dependence on $b$ in much of our notation.

\begin{figure}[htb]
        \centering
        \includegraphics[width=0.8\linewidth]{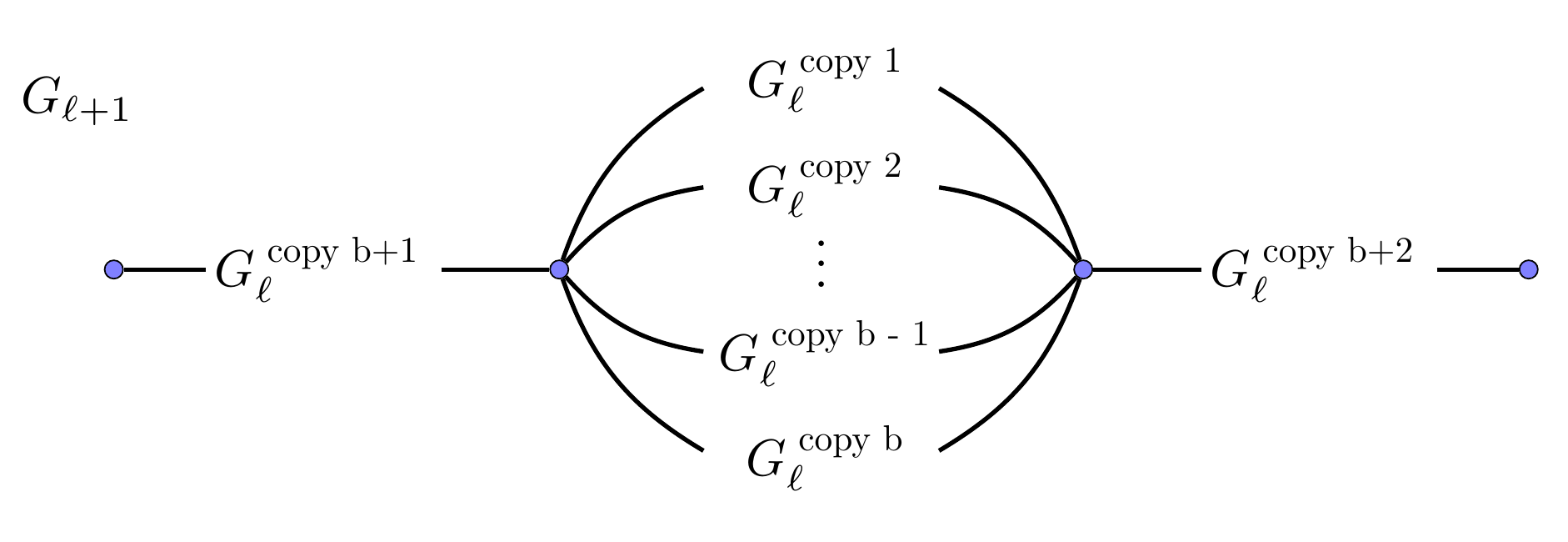}
    \caption{$G_{\ell+1}$ is constructed by gluing $b+2$ copies  of $G_{\ell}$, labeled as shown.}
    \label{fig:copiesLabel}
\end{figure}

The construction of $G_{\ell+1}$ from Definition~\ref{def:BubbleDiamondGraphs} is equivalent to replacing each edge of $G_\ell$ with a copy of $G_1$. Alternatively, one can make the construction by replacing the edges of $G_1$ by copies of $G_\ell$.  The latter is useful for defining a limit of the sequence $G_\ell$, as we may identify $G_\ell$ with one of these copies and then set $G_\infty=\cup G_\ell$. However the result depends on the choices of identifications in the sequence. To keep track of them we label the central copies of $G_\ell$ in $G_{\ell+1}$ by  $\{1, \dots, b\}$ and the left and right side copies by $b+1$ and $b+2$ as in Figure~\ref{fig:copiesLabel} and make the following definition.

\begin{definition}
\label{def:infiniteBubbleDiamond}
Let $\{k_{\ell}\}_{\ell = 0}^{\infty}$ be a sequence such that $k_{\ell} \in \{1,2,\dots , b+2\}$. Consider the bubble diamond graphs $\{G_{\ell}\}_{\ell \geq 0}$ as an increasing sequence in which $G_\ell$ is identified with the $k_\ell$ copy in $G_{\ell+1}$ according to the labeling in Figure~\ref{fig:copiesLabel}. The corresponding  \textit{infinite Bubble diamond graph} is $G_{\infty} = \cup_{\ell =0}^{\infty} G_{\ell}$, meaning that  $V(G_{\infty}) = \cup_{\ell =0}^{\infty} V(G_{\ell})$ and $E(G_{\infty}) = \cup_{\ell =0}^{\infty} E(G_{\ell})$.
\end{definition}
It is not difficult to see that there are uncountably many non-isomorphic limits $G_\infty$ and that any $G_\infty$ is locally finite, meaning that any vertex has finite degree.

We define a Hilbert space $L^2(G_{\infty})$ of complex-valued functions on the vertices of $G_\infty$ with weights from the degree, so that the inner product is $\bra{f} \ket{g} = \sum_{p \in V(G_{\infty})} f(p) \overline{g(p)} \deg(p)$. 
The probabilistic graph Laplacian of a function $f \in L^2(G_{\infty})$ is then defined to be
\begin{equation}
\label{eq:probabilisticGraphLaplacianinfinite}
    \Delta_{\infty,b} f(q)= f(q)-\frac{1}{\deg(q)}\sum_{(q,p)\in E(G_{\infty})} f(p), \quad q \in V(G_{\infty}).
\end{equation}
and is a bounded self-adjoint operator on $L^2(G_{\infty})$ with spectrum $\sigma(\Delta_{\infty,b}) \subset [0,2]$. In a similar manner we consider $L^2(G_{\ell})$ with inner product $\bra{f} \ket{g} = \sum_{p \in V(G_{\ell})} f(p) \overline{g(p)} \deg_{\ell}(p)$ and  graph Laplacian 
\begin{equation}
\label{eq:probabilisticGraphLaplacian}
    \Delta_{\ell,b} f(q)= f(q)-\frac{1}{\deg_{\ell}(q)}\sum_{(q,p)\in E(G_{\ell})} f(p), \quad \ q \in V(G_{\ell}).
\end{equation}

For $\ell \geq 1$, the Dirichlet graph Laplacian $\Delta^D_{\ell,b}$ is defined by \eqref{eq:probabilisticGraphLaplacian} but on the domain 
\begin{equation}
\label{eq:domain}
\{ f \in  L^2(G_{\ell}) \ : \  f|_{\partial G_{\ell}}=0 \ \}. 
\end{equation}

\begin{theorem}
\label{thm:purepointSpectrum}
Fix a sequence $\{k_{\ell}\}_{\ell = 0}^{\infty}$,  $k_{\ell} \in \{1,2,\dots , b+2\}$  for which $k_{\ell} \in \{1,2,\dots , b\}$ infinitely often. Let $G_{\infty}$ be the corresponding infinite Bubble diamond graph and $\Delta_{\infty,b}$ be the associated probabilistic graph Laplacian. Then the spectrum of $\Delta_{\infty,b}$ is pure point and given by 
\begin{equation}
\sigma(\Delta_{\infty,b}) = \overline{\cup_{\ell = 1}^{\infty} \sigma( \Delta^D_{\ell,b}) }.
\end{equation}
Moreover, there is a complete set of 
compactly supported eigenfunctions and every eigenvalue is of infinite multiplicity.
\end{theorem}
\begin{remark}
Theorem~\ref{thm:purepointSpectrum} is almost a special case of Theorem~5.8 in~\cite{MalozemovTeplyaev2003}, but unfortunately that result cannot be directly applied because $G_1$ is not a $2$-point model graph in the sense of \cite[Definition 5.1.]{MalozemovTeplyaev2003}. Although the proof requires only minor changes from that in~\cite{MalozemovTeplyaev2003} we provide the details below.
\end{remark}

\begin{figure}[htb]
        \centering
        \includegraphics[width=0.85\linewidth]{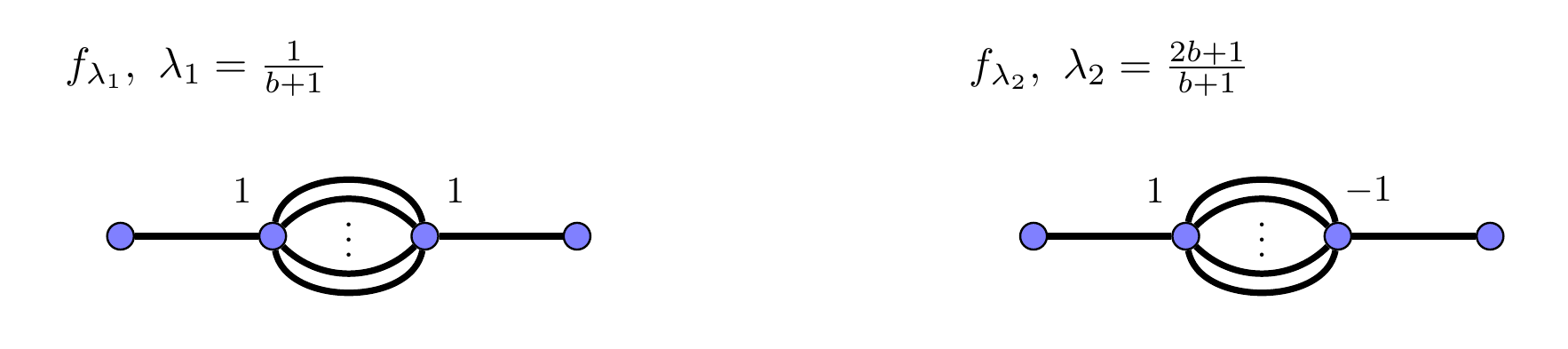}
    	  \caption{Direct computation gives $\sigma(\Delta^D_{1,b})= \{   \frac{1}{b + 1}, \  \frac{2 b + 1}{b + 1} \}$ and corresponding eigenfunctions $f_{\lambda_1}$ and $f_{\lambda_1}$. Values of the eigenfunctions are shown on vertices where they are non-zero.}
    	  \label{fig:antisymmetric1Level1}
\end{figure}

\begin{figure}[htb]
        \centering
        \includegraphics[width=0.8\linewidth]{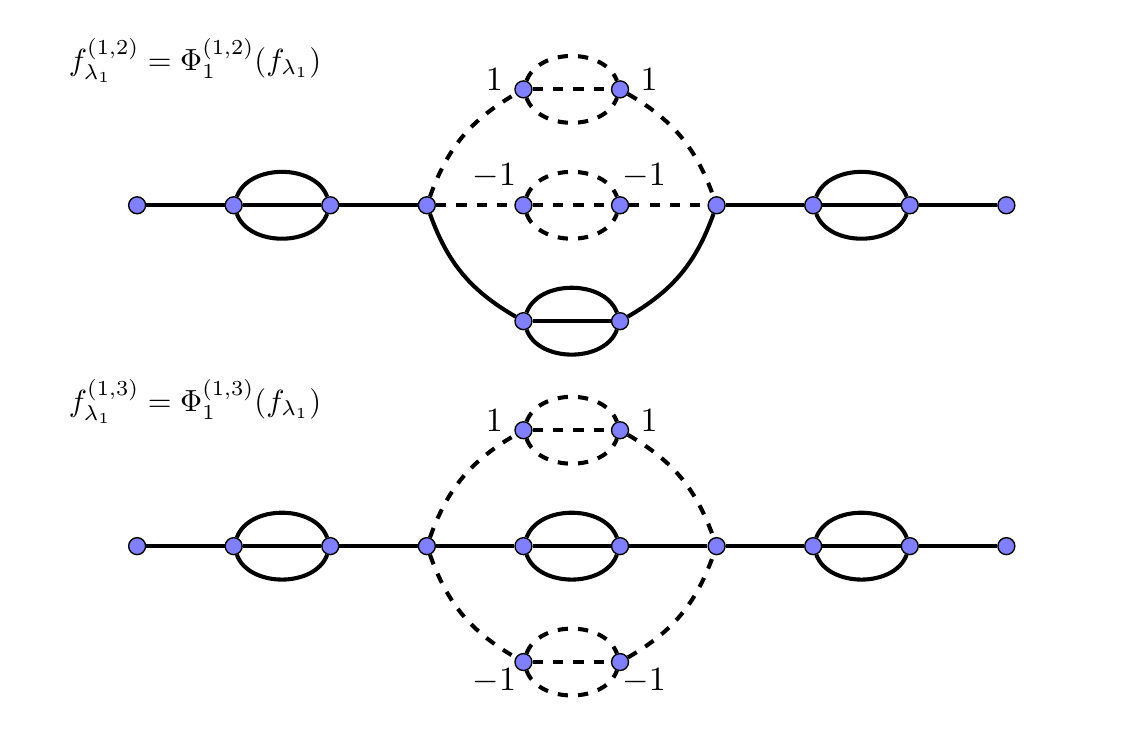}
    	  \caption{The eigenfunctions $\Phi_{1}^{(1,2)}(f_{\lambda_1})$ and $ \Phi_{1}^{(1,3)}(f_{\lambda_1})$ are supported on the dashed edges. Values of the eigenfunctions at vertices are included where they are non-zero.}
    	  \label{fig:antisymmetric1Level2constEig}
\end{figure}

The proof relies on an elementary construction of eigenfunctions of $\Delta_{\infty,b}$ from eigenfunctions of $\Delta^D_{\ell,b}$ via an operator that is defined for $\ell\geq1$ and $i,j\in \{1,2,\dots , b\}$ with $i\neq j$ by
\begin{align}
\Phi_{\ell}^{(i,j)}: \{ f \in  L^2(G_{\ell}) \ &: \  f|_{\partial G_{\ell}}=0 \ \} \to   L^2(G_{\infty})  \quad \nonumber
\\
\Phi_{\ell}^{(i,j)}(f) &=
  \begin{cases}
    f       &  \text{on the $i$th copy of $G_{\ell}$ in $G_{\ell+1}$},\\
    -f       &  \text{on the $j$th copy of $G_{\ell}$ in $G_{\ell+1}$},\\
    0          &     \text{ elsewhere in $G_{\infty}$}.
  \end{cases} 
\end{align}
The Dirichlet condition ensures this is well-defined, it is evident that $||\Phi_{\ell}^{(i,j)}(f)|| \leq 2||f||$, and the following lemma is easily checked. A similar idea was used in~\cite{MalozemovTeplyaev1995}.
\begin{lemma} \label{lem:ConstructofEigenfunc}
If $ f_{\lambda} \in  L^2(G_{\ell})$ is an eigenfunction of $\Delta^D_{\ell,b}$ with eigenvalue $\lambda$ then $f_{\lambda}^{(i,j)}:=\Phi_{\ell}^{(i,j)}(f_{\lambda})$  is an eigenfunction of $\Delta_{\infty,b}$ with the same eigenvalue. Moreover, $f_{\lambda}^{(i,j)}$ has finite support $\text{supp}(f_{\lambda}^{(i,j)}) \subset V(G_{\ell+1})$ and its restriction to $G_{\ell+1}$ is an eigenfunction of both $\Delta_{\ell+1,b}$ and $\Delta^D_{\ell+1,b}$.
\end{lemma}
Figure~\ref{fig:antisymmetric1Level1} shows two Dirichlet eigenfunctions on $G_1$, and Figure~\ref{fig:antisymmetric1Level2constEig} illustrates the construction of Lemma~\ref{lem:ConstructofEigenfunc} applied to the first of these eigenfunctions.

\begin{proof}[Proof of Theorem \ref{thm:purepointSpectrum}]
Define a subsequence $\{k_{\ell_n}\}_{n = 1}^{\infty}$ of  $\{k_{\ell}\}_{\ell = 1}^{\infty}$  by including only those $k_{\ell} \in \{1,2,\dots , b\}$. Suppose that for each $\ell_i$ we have fixed a maximal linearly independent set $\Lambda_{\ell_i}$ of eigenfunctions of $\Delta^D_{\ell_i,b}$ and define
\begin{equation}
S_{n}= \bigcup_{i=1}^{n-1} \Big \{ \Phi_{\ell_i}^{(k_{\ell_i},j)}(f)  \ \Big| \ j \in  \{1,\dots , b\}\backslash \{k_{\ell_i}\}, \text{ and } f\in\Lambda_{\ell_i} \ \Big \}
\end{equation}
and then $S= \cup_{n=2}^{\infty} S_{n}$. By Lemma~\ref{lem:ConstructofEigenfunc} $S$ is a set of compactly supported eigenfunctions of $\Delta_{\infty,b}$ and it is obvious each eigenvalue has infinite multiplicity.  We show $S$ is complete.

Let $f \in L^2(G_{\infty})$ be orthogonal to the span of $S$. Define $f_{\ell_n}$ to be the projection to $G_{\ell_n}\setminus \partial G_{\ell_n}$, so $f_{\ell_n}=f$ on this set and is zero elsewhere and observe that $\|f-f_{\ell_n}\|\to0$ as $n\to\infty$.   Identifying $f_{\ell_n}$ with a function in $L^2(G_{\ell_n})$ we see that it vanishes on $\partial G_{\ell_n}$ and is  therefore in the span of the eigenfunctions of $\Delta^D_{\ell_n,b}$.  It follows from the definition of $S$ that if $j\in\{1,\dotsc, b\}\setminus\{k_{\ell_n}\}$ then $\Phi_{\ell_n}^{(k_{\ell_n},j)}(f_{\ell_n})$ is in the span of $S$ and is therefore orthogonal to $f$ by hypothesis. However $f=f_{\ell_n}=\Phi_{\ell_n}^{(k_{\ell_n},j)}(f_{\ell_n})$ on $G_{\ell_n}\setminus\partial G_{\ell_n}$, from which
\begin{equation*}
	0=\bra{f}\ket{\Phi_{\ell_n}^{(k_{\ell_n},j)}(f_{\ell_n})}
	=\|f_{\ell_n}\|^2 + \bra{f-f_{\ell_n}} \ket{\Phi_{\ell_n}^{(k_{\ell_n},j)}(f_{\ell_n})}
	\end{equation*}
Using the fact that the part of $\Phi_{\ell_n}^{(k_{\ell_n},j)}(f_{\ell_n})$ on the complement of $G_{\ell_n}\setminus \partial G_{\ell_n}$ is a copy of $-f_{\ell_n}$ we thereby deduce $\|f_{\ell_n}\|^2\leq \|f-f_{\ell_n}\|\|f_{\ell_n}\|\leq  \|f-f_{\ell_n}\|\|f\|$. Since $\|f-f_{\ell_n}\|\to0$ this proves $f=\lim_n f_{\ell_n}=0$, which completes the proof. 
\end{proof}

It is significant that although the spectrum as a set does not depend on the choice of the sequence $k_{\ell}$, see Theorem~\ref{thm:spectrumOfinfiniteLap},  the type of the spectrum is dependent on the choice of $k_{\ell}$. To illustrate the interesting subtleties of this family of self-similar graphs, we contrast the following result with Theorem~\ref{thm:purepointSpectrum}.
\begin{theorem}\label{prop:sc} Assume that either $k_{\ell} =b+1$ for all $l\geqslant0$, or that $k_{\ell} =b+2$ for all $l\geqslant0$.   
Then the spectrum of $\Delta_{\infty,b}$ has a pure point component and a singularly continuous component with supports equal to  $\mathcal{J}(R_b)$.
\end{theorem}
\begin{proof}
For this choice of $k_{\ell}$, the infinite Bubble diamond graph $G_{\infty}$ is a $\integers_+$-graded graph in the sense of \cite{SolitonsGradedGraphs2021Mograby}. The statement follows from the proofs in Theorem~\ref{thm:purepointSpectrum} and in \cite[Theorem 1]{ChenTeplyaev2016}, as in this context $\Delta_{\infty,b}$ can be represented as a lift of $\Delta_{p}$ (the $pq$-model in \cite{ChenTeplyaev2016}). This is done for $p=\frac{b}{b+1}$ using
\cite[Corollary 2.15]{SolitonsGradedGraphs2021Mograby}.
\end{proof}

\section{Spectral decimation for Bubble-diamond graphs}
\label{sec-spectralDeci}
We briefly review a technique common in Analysis on Fractals called \textit{Spectral Decimation}. Its central idea is that the spectrum of a Laplacian on fractals or self-similar graphs built from pieces that satisfy some strong symmetry assumptions can be completely described in terms of iterations of a rational function called the \textit{spectral decimation function}. We show in this section that the spectral decimation function for the Bubble-diamond graphs is a polynomial and prove the following result. Our arguments rely heavily on ideas and results from~\cite{MalozemovTeplyaev2003}.
\begin{theorem}
\label{thm:spectrumOfinfiniteLap}
$\sigma(\Delta_{\infty,b})=\mathcal{J}(R_b)$, where $\mathcal{J}(R_b)$ is the Julia set of the polynomial $R_b$ given in (\ref{eq:spectralDeciFunction}).
\end{theorem}

\begin{definition}[\protect{\cite[Definition 2.1]{MalozemovTeplyaev2003}}]
\label{def:specSimiResol}
Let $\mathcal{H}$ and $\mathcal{H}_0$ be Hilbert spaces, and $U:\mathcal{H}_0 \to \mathcal{H}$ be an isometry. Suppose  $H$ and $H_0$ are bounded linear operators on  $\mathcal{H}$ and $\mathcal{H}_0$, respectively, and that $\phi,\psi$ are complex-valued functions. We call the operator $H$ \textit{spectrally similar} to the operator $H_0$ with functions $\phi$ and $\psi$ if
\begin{equation}
\label{eq:OriginalSpectSimi}
U^{\ast}(H-z)^{-1}U=(\phi(z)H_0 - \psi(z))^{-1},
\end{equation}
for all $z \in \complex$ such that the two sides of~\eqref{eq:OriginalSpectSimi} are well defined. Note, in particular, that for $z$ in the domain of both $\phi$ and $\psi$ and satisfying $\phi(z)\neq0$ we have $z\in\rho(H)$ (the resolvent set) if and only if $R(z)=\frac{\psi(z)}{\phi(z)}\in\rho(H_0)$.  We call $R(z)$ the {\em spectral decimation function}.
\end{definition}
The functions $\phi(z)$ and $\psi(z)$ are difficult to read directly from the structure of the considered fractal or graph, but they can be computed effectively using a Schur complement (several examples may be found in~\cite{MalozemovTeplyaev2003,BajorinVibration3Ngasket2008,BajorinVibrationSpectra2008}). Identifying $\mathcal{H}_0$  with a closed subspace of $\mathcal{H}$ via $U$, let $\mathcal{H}_1$ be the orthogonal complement and decompose $H$ on   $\mathcal{H}=\mathcal{H}_0 \oplus \mathcal{H}_1$ in the block form
\begin{equation}
\label{eq:lDecompo}
H=\begin{pmatrix}
T & J^T\\
J & X
\end{pmatrix}.
\end{equation}
\begin{lemma}[\protect{\cite{MalozemovTeplyaev2003}, Lemma 3.3}]
\label{lem:Lemma33}
For $z \in \rho(H)\cap \rho(X)$ the operators $H$ and $H_0$ are spectrally similar if and only if
the Schur complement of $H-zI$, given by $S_{H}(z)=T-z -  J^T(X-z)^{-1}J$, satisfies
\begin{equation}
\label{eq:schurComp1}
    S_{H}(z)=  \phi(z) H_0 - \psi(z)I.
\end{equation}
\end{lemma}
The set  $\mathscr{E}_{H}:=\{z \in \complex \ | \ z \in \sigma(X) \text{ or } \phi(z)=0 \}$ plays a crucial role in the spectral decimation method and we refer to it as the \textit{exceptional set} of $H$.  The following could be derived immediately from Lemma~4.2 of~\cite{MalozemovTeplyaev2003}, but it is convenient to calculate it directly so as to obtain explicit formulas for $\phi_b$ and $\psi_b$.

\begin{corollary}\label{cor:specdeclevel1}
$\Delta_{1,b}$ is spectrally similar to $\Delta_{0,b}$.
\end{corollary}
\begin{proof}
We have $H=L^2(G_1)$ and $H_0=L^2(G_0)$ and may directly compute from~\eqref{eq:probabilisticGraphLaplacian} that
\begin{equation*}
	\Delta_{1,b} = \frac1{b+1}
				\begin{pmatrix}
				b+1&0& -(b+1)&0 \\
				 0&b+1&0&-(b+1) \\
				-1&0&b+1&-b\\
				0& -1& -b& b+1
				 \end{pmatrix}, \quad
	\Delta_{0,b} = \begin{pmatrix} 1&-1\\-1&1\end{pmatrix}.
	\end{equation*}
The Schur complement is found to be $S_{\Delta_{1,b}}= \phi(z) H_0 - \psi(z)I$ where the functions are
\begin{equation}\label{eq:specDeciFuncLap}
\phi_b(z) = \frac{b}{  \left(b + 1\right)^{2} \left(z - 1\right)^{2} - b^{2}} , \quad \quad \psi_b(z)  = \frac{z ((b+1) z - b  - 2)}{(b+1) z - 1}.
	\end{equation}
The exceptional set $\mathscr{E}_{\Delta_{\ell,b}}$ is therefore $\mathscr{E}_{b} = \left\{ \frac{2 b + 1}{b + 1}, \  \frac{1}{b + 1}\right\}$.
\end{proof}

The significance of Corollary~\ref{cor:specdeclevel1} comes from the fact that graphs built from copies of spectrally equivalent pieces are themselves spectrally equivalent. A precise version of this is found in Lemma~3.10 of~\cite{MalozemovTeplyaev2003}, but since we are considering the probabilistic Laplacians $\Delta_{l,b}$ we can apply the more convenient  Lemma~4.7 of~\cite{MalozemovTeplyaev2003}, viewing the graph $G_l$ as having been obtained from $G_{l-1}$ by replacing copies of $G_0$ with $G_1$ to find spectral similarity of both Neumann and Dirichlet Laplacians at every level. We record this as a proposition.

\begin{proposition}
 \label{prop: SpectralDecimpLap}
Let $\ell \geq 2$. Then $\Delta_{\ell-1,b}$ is spectrally similar to $\Delta_{\ell-2,b}$ and $\Delta^D_{\ell,b}$ is spectrally similar to $\Delta^D_{\ell-1,b}$, in both cases with respect to the functions
 $\phi_b$ and $\psi_b$ in~\eqref{eq:specDeciFuncLap}. The exceptional set is $\mathscr{E}_{b} = \left\{ \frac{2 b + 1}{b + 1}, \  \frac{1}{b + 1}\right\}$ and the spectral decimation function is the third order polynomial 
\begin{equation}
\label{eq:spectralDeciFunction}
R_b(z)= \frac{\psi_b(z)}{\phi_b(z)}= \frac zb  \bigl( (b+1) z - 2 b - 1\bigr)\bigl((b+1) z - b  - 2\bigr).
\end{equation}
\end{proposition}

In Definition~\ref{def:specSimiResol} we noted a relation between the resolvents of spectrally similar operators, which in our case is now seen to imply that $z\in\sigma(\Delta_{l,b})\setminus\mathscr{E}_{b}$ if and only if $R_b(z)\in\sigma(\Delta_{l-1,b})$, and similarly for the Dirichlet Laplacians.  Moreover, the spectral similarity induces a bijection between the corresponding eigenspaces so the multiplicity of $z$ is the same as that of $R_b(z)$, see Theorem~3.6 of~\cite{MalozemovTeplyaev2003}.  It is not difficult to use these observations to count the eigenvalues $\sigma(\Delta_{l,b})$ and their multiplicities in  a manner similar to that used in~\cite{BajorinVibration3Ngasket2008}; we do the latter only for the Dirichlet case because it will be needed later.  The following easily verified properties of $R_b$ will be useful.

\begin{lemma}\label{lem:dynamofRb}
The fixed points of $R_b$ are $\{0,1,2\}$, all of which are repelling. There are critical points in each of $(0,1)$ and $(1,2)$ and the critical values are outside $[0,2]$, so every point in $[0,2]$ has three distinct preimages under $R_b^{-1}$. Also
\begin{equation}\label{eq:preimsof02}
R_b^{-1} (0)=\Big \{0, \  \frac{b + 2}{b + 1}, \  \frac{2 b + 1}{b + 1} \Big \}, \quad \quad R_b^{-1} (2 )=\Big \{  \frac{1}{b + 1}, \ \frac{b}{b + 1} , \ 2 \Big \}.
\end{equation}
In particular, $R_b^{-1}(z)\cap\mathscr{E}_{b}=\emptyset$ unless $z\in\{0,2\}$.
\end{lemma}
\begin{proof}
The fixed points can easily be verified, as can the sets in~\eqref{eq:preimsof02}. Direct computation of the derivative shows it is $(2b+1)(b+1)/b>1$ at both $0$ and $2$ and $-(b^2+b+1)/b<-1$ at $1$. The observation that $R_b^{-1}(0)$ contains two distinct points in $(1,2)$ and $R_b^{-1}(2)$ contains two distinct points in $(0,1)$ implies both that each interval contains a critical point and that the critical values are outside $[0,2]$. The last statement follows from the fact that $0$ and $2$ are fixed points and~\eqref{eq:preimsof02}.
\end{proof}

\begin{proposition}
\label{prop:SpectralNeumannFinite}
Let $b \geq 2$ and $\{\Delta_{\ell,b}\}_{\ell \geq 0}$ be the Laplacians from~\eqref{eq:probabilisticGraphLaplacian}.  Then
$\sigma(\Delta_{0})=\{0,2\}$, $\sigma(\Delta_{1,b}) = \Big\{ 0, \frac{b }{b + 1}  ,  \frac{b + 2}{b + 1} , 2 \Big\}$ and for $\ell\geq2$, $\sigma(\Delta_{\ell,b}) =R_b^{-\ell}(\{0,2\})$.
\end{proposition}

\begin{proof}
We compute $\Delta_{0,b}=\{0,2\}$ and observe that each eigenvalue has multiplicity $1$. The discussion about spectral decimation that precedes Lemma~\ref{lem:dynamofRb} ensures that for all $\ell\geq1$ 
\begin{equation}\label{eq:spectdecinclusion}
	R_b^{-1}\bigl(\sigma(\Delta_{\ell-1,b})\bigr)\setminus\mathscr{E}_{b}
	\subset \sigma(\Delta_{\ell,b})
	\subset R_b^{-1}\bigl(\sigma(\Delta_{\ell-1,b})\bigr)\cup\mathscr{E}_{b}.
	\end{equation}
In particular $R_b^{-1}(\{0,2\})\setminus\mathscr{E}_{b}\subset\sigma(\Delta_{1,b})$ with each eigenvalue having multiplicity $1$. Comparing this to~\eqref{eq:preimsof02} and using the fact that $|V(G_1)|=4$ we obtain the stated formula for $\sigma(\Delta_{1,b})$.

We also see from~\eqref{eq:preimsof02} that $\mathscr{E}_{b}\subset R_b^{-1}(\{0,2\})$, and since another application of~\eqref{eq:spectdecinclusion} gives $\{0,2\}\subset \sigma(\Delta_{\ell,b})$ we can rewrite~\eqref{eq:spectdecinclusion} as
\begin{equation*}
	R_b^{-1}\bigl(\sigma(\Delta_{\ell-1,b})\bigr)\setminus\mathscr{E}_{b}
	\subset \sigma(\Delta_{\ell,b})
	\subset R_b^{-1}\bigl(\sigma(\Delta_{\ell-1,b})\bigr).
	\end{equation*}
We recall  that for each $z\in\mathscr{E}_{b}$ there is an eigenfunction of $\Delta^D_{1,b}$ with eigenvalue $z$; this was illustrated in Figure~\ref{fig:antisymmetric1Level1}.  Inductively applying Lemma~\ref{lem:ConstructofEigenfunc}, which says that an eigenvalue of $\Delta^D_{l,b}$ is also an eigenvalue of both $\Delta^D_{l+1,b}$ and $\Delta_{l+1,b}$ we find that $\mathscr{E}_{b}\subset \sigma(\Delta_{\ell,b})$ once $\ell\geq2$.
\end{proof}

\begin{proposition}
\label{prop:SpectralDiriFinite}
Let $b \geq 2$ and $\{\Delta^D_{\ell,b}\}_{\ell \geq 1}$ be the sequence of Dirichlet graph Laplacians given in (\ref{eq:probabilisticGraphLaplacian}) with the domain (\ref{eq:domain}).  Then for $\ell \geq 1$: 
\begin{enumerate}
\item $\sigma(\Delta^D_{\ell,b})=\bigcup_{n=0} ^{\ell-1} R_b^{-n} (\mathscr{E}_b) \subset \sigma(\Delta_{\ell+1,b})$
\item If $ 0\leq m \leq \ell$ and $z\in R_b^{-m}(\mathscr{E}_b)$ then $\mult_{\ell+1} (z)=\frac{(b-1)(b+2)^{\ell-m} +2}{b+1}$.
\end{enumerate}
\end{proposition}
\begin{proof}
The proof is much like that for the previous lemma.
We have computed $\sigma(\Delta^D_{1,b})=\mathscr{E}_{b}$; see also Figure~\ref{fig:antisymmetric1Level1}. As noted in Lemma~\ref{lem:dynamofRb},  $R_b^{-\ell}(\sigma(\Delta^D_{1,b}))\cap\mathscr{E}_{b}=\emptyset$ if $\ell\geq1$, so by the  discussion about spectral decimation that precedes Lemma~\ref{lem:dynamofRb} we find for $\ell\geq1$ that
\begin{equation}\label{eq:spectdecforDir}
	R_b^{-1}\bigl(\sigma(\Delta^D_{\ell,b})\bigr)
	\subset \sigma(\Delta_{\ell+1,b})
	\subset R_b^{-1}\bigl(\sigma(\Delta_{\ell-1,b})\bigr)\cup\mathscr{E}_{b}.
	\end{equation}
From Lemma~\ref{lem:ConstructofEigenfunc} we have $\sigma(\Delta^D_{\ell,b})\subset\sigma(\Delta^D_{\ell+1,b})$, from which $\mathscr{E}_{b}=\sigma(\Delta^D_{1,b})\subset\sigma(\Delta^D_{\ell,b})$ for all $\ell\geq1$. Combining this and~\eqref{eq:spectdecforDir} gives $\sigma(\Delta^D_{\ell+1,b})=\cup_0^\ell R_b^{-n}(\mathscr{E}_{b})$ as stated. The second inclusion in~(1) is from Lemma~\ref{lem:ConstructofEigenfunc}.

Take $z\in\sigma(\Delta^D_{\ell+1,b})$; the preceding says there is $0\leq n\leq \ell$ so $R_b^n(z)\in\mathscr{E}_b$. By the discussion  preceding Lemma~\ref{lem:dynamofRb}, $\mult_{l+1}(z)=\mult_l(R_b(z))$ unless $z\in\mathscr{E}_{b}$, and iterating gives $\mult_{\ell+1}(z)=\mult_{\ell+1-n}(R_b^n(z))$. Accordingly we can determine all multiplicities by determining those for $w\in\mathscr{E}_{b}$ at all scales.

Since there are three preimages of each point of $\sigma(\Delta^D_{k,b})$ under $R_b$ the number of eigenvalues, counting multiplicity, of $\Delta^D_{k+1,b}$ obtained in this fashion is $3(|V(G_k)|-2)$.  Accordingly, the total multiplicity corresponding to $\mathscr{E}_b$ is $|V(G_{k+1})|-3|V(G_k)|+4$. It is easy to compute $|V(G_k)|=\frac2{b+1}\bigl((b+2)^k+b\bigr)$, so the total multiplicity for $\Delta^D_{k+1,b}$ of the two eigenvalues in $\mathscr{E}_b$ is $\frac{2}{b+1}\bigl((b-1)(b+2)^k+2\bigr)$.

It remains to be seen that both eigenvalues in $\mathscr{E}$ have the same multiplicity at each level. Recall that Lemma~\ref{lem:ConstructofEigenfunc} took an eigenfunction $f$ of $\Delta^D_{\ell,b}$ and constructed $b-1$ linearly independent eigenfunctions of $\Delta^D_{\ell+1b}$, all supported on the central $b$ copies of $G_{\ell-1}$ in $G_{\ell}$. These were also eigenfunctions of $\Delta_{\ell+1,b}$, so we call them Dirichlet-Neumann eigenfunctions and write $\mult^{DN}$ for their multiplicity. Thus we have shown $\mult^{DN}_{\ell+1}(w)\geq(b-1)\mult_{\ell}(w)$.  However, if $f$ is a Dirichlet-Neumann eigenfunction on $G_l$ with eigenvalue $w$ then it is obvious that placing a copy of $f$ on any of the $(b+2)$ copies of $G_{\ell}$ in $G_{\ell+1}$ shown in Figure~\ref{fig:copiesLabel}, and extending by zero to the rest of $G_{\ell+1}$, gives a Dirichlet-Neumann eigenfunction with eigenvalue $w$. These eigenfunctions are linearly independent, so $\mult^{DN}_{\ell+1}(w)\geq(b+2)\mult^{DN}_{\ell}$. Combining this with the preceding gives
\begin{equation}\label{eq:DNcounting1}
	\mult^{DN}_{\ell+1}(w)\geq (b+2)\mult^{DN}_{\ell}+(b-1)(\mult_{\ell}(w)-\mult^{DN}_{\ell}(w)).
	\end{equation}
We make one further construction. Given a $\Delta^D_{\ell,b}$ eigenfunction $f$ which is not Dirichlet-Neumann we can place a copy of a multiple of $f$ on each of the copies $b+1,1,b+2$ of $G_l$ in $G_{\ell+1}$ following the labelling in Figure~\ref{fig:copiesLabel}.  Choosing the coefficients of these multiples so that the graph Laplacian is zero at the two vertices where these copies meet, we see this defines an eigenfunction of $\Delta^D_{\ell+1,b}$ that is linearly independent of the $b-1$ constructed by the method of Lemma~\ref{lem:ConstructofEigenfunc}. This eigenfunction is not Dirichlet-Neumann. Thus $\mult_{\ell+1}(w)-\mult^{DN}_{\ell+1}(w)\geq \mult_{\ell}(w)-\mult^{DN}_{\ell}(w)$, and since $\mult_1(w)-\mult^{DN}_1(w)=1$ for the $w\in\mathscr{E}_b$ we get $\mult_{\ell+1}(w)-\mult^{DN}_{\ell+1}(w)\geq1$ for all $\ell$ in this case.  From~\eqref{eq:DNcounting1} then
$\mult^{DN}_{\ell+1}(w)\geq (b+2)\mult^{DN}_{\ell}(w)+(b-1)$ and inductively, beginning with $\mult^{DN}_2(w)=b-1$, we have $\mult^{DN}_{\ell+1}(w)\geq \frac{(b-1)}{b+1}\bigl((b+2)^{\ell}-1\bigr)$.  Since we also have $\mult_{\ell+1}(w)-\mult^{DN}_{\ell+1}(w)\geq1$ we obtain $\mult_{\ell+1}(w)\geq \frac1{b+1}\bigl((b-1)(b+2)^{\ell}+2\bigr)$. This applies to each $w\in\mathscr{E}_b$, but the converse bound on the sum of their multiplicities was established above, so equality must hold.

The above reasoning gives $\mult_{\ell+1}(z)=\mult_{\ell+1-n}(R_b^n(z))=\frac1{b+1}\bigl((b-1)(b+2)^{\ell-n}+2\bigr)$
for $z\in R_b^{-n}(\mathscr{E}_b)$.
\end{proof}

\begin{proof}[Proof of Theorem \ref{thm:spectrumOfinfiniteLap}]
Recall that the Julia set $\mathcal{J}(R_b)$ is the closure of the backward orbit of any of its elements, and that it contains all repelling fixed points. A suitable reference is~\cite{MR1230383} or Chapter~14 of~\cite{FalconerBookFractGeo2003}.

We saw in Lemma~\ref{lem:dynamofRb} that $R_b(\mathscr{E}_b)=\{0,2\}$ and that these latter are repelling fixed points, so $\mathscr{E}_b\subset\mathcal{J}(R_b)$.  Then from Theorem~\ref{thm:purepointSpectrum} and Proposition~\ref{prop:SpectralDiriFinite} we have
\begin{equation*}
	\sigma(\Delta_{\infty,b})
	=\overline{\cup_{\ell=1}^\infty \sigma\bigl(\Delta^D_{\ell,b}\bigr)}
	= \overline{\cup_{\ell=1}^\infty R_b^{-\ell}(\mathscr{E}_b)}
	=\mathcal{J}(R_b)
	\end{equation*}
where the last equality holds because $\mathcal{J}(R_b)$ is the closure of the backward orbit of any point from $\mathcal{J}(R_b)$.
\end{proof}

%
%
%
%
%
%
%
%

\section{The Integrated Density of States of $\Delta_{\infty,b}$}\label{sec-IDS}
\begin{figure} 
  \begin{minipage}[b]{0.5\linewidth}
    \centering
    \includegraphics[width=1.0\linewidth]{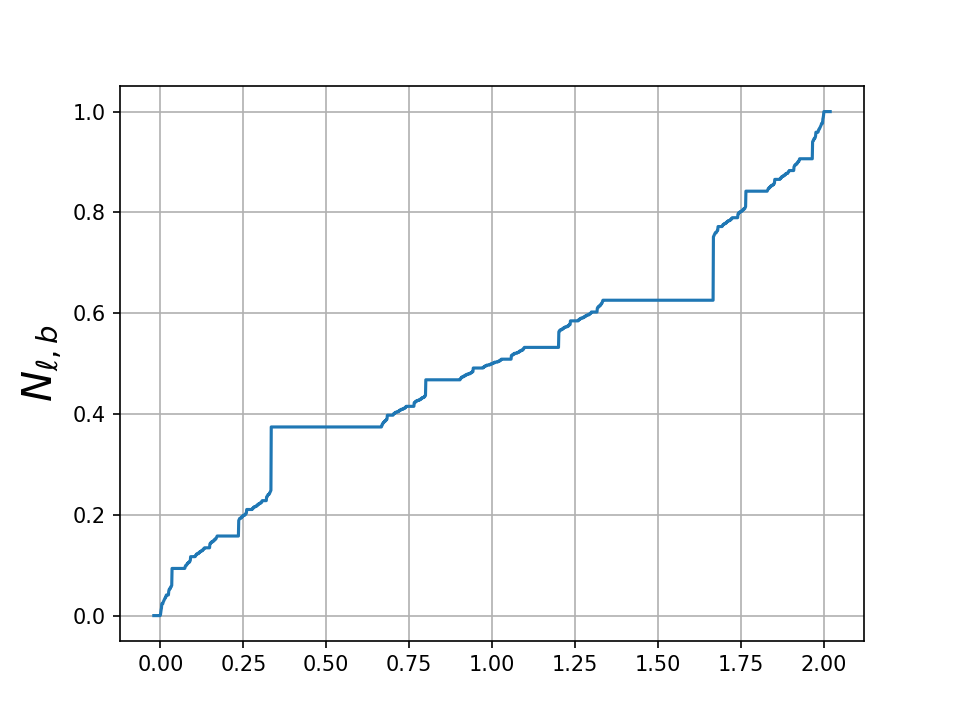} 
  \end{minipage}
  \begin{minipage}[b]{0.5\linewidth}
    \centering
    \includegraphics[width=1.0\linewidth]{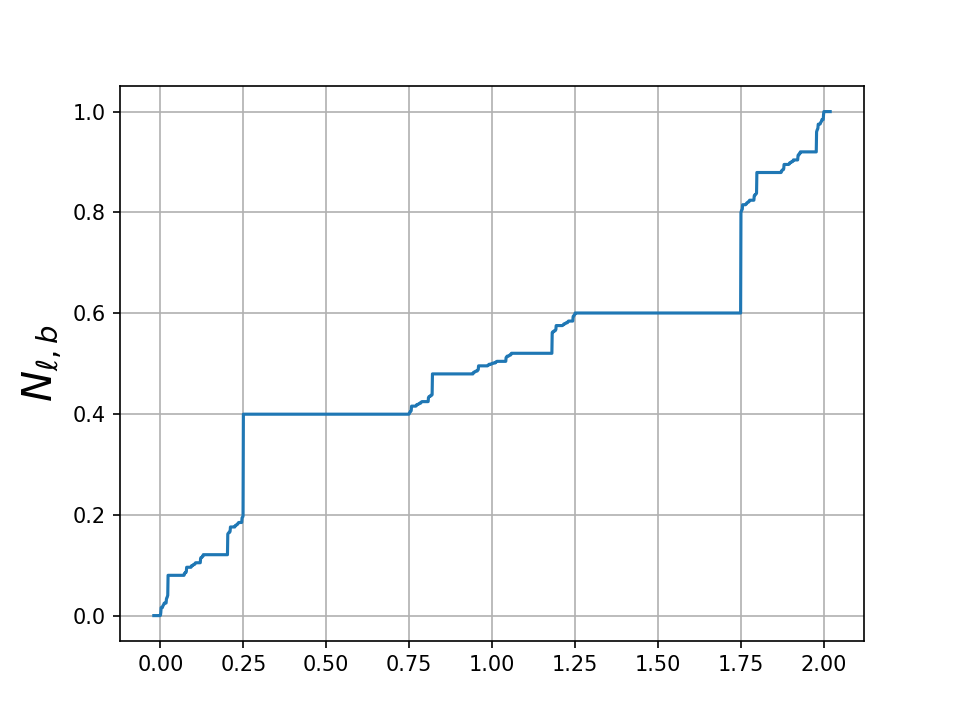} 
  \end{minipage} 
  \begin{minipage}[b]{0.5\linewidth}
    \centering
    \includegraphics[width=1.0\linewidth]{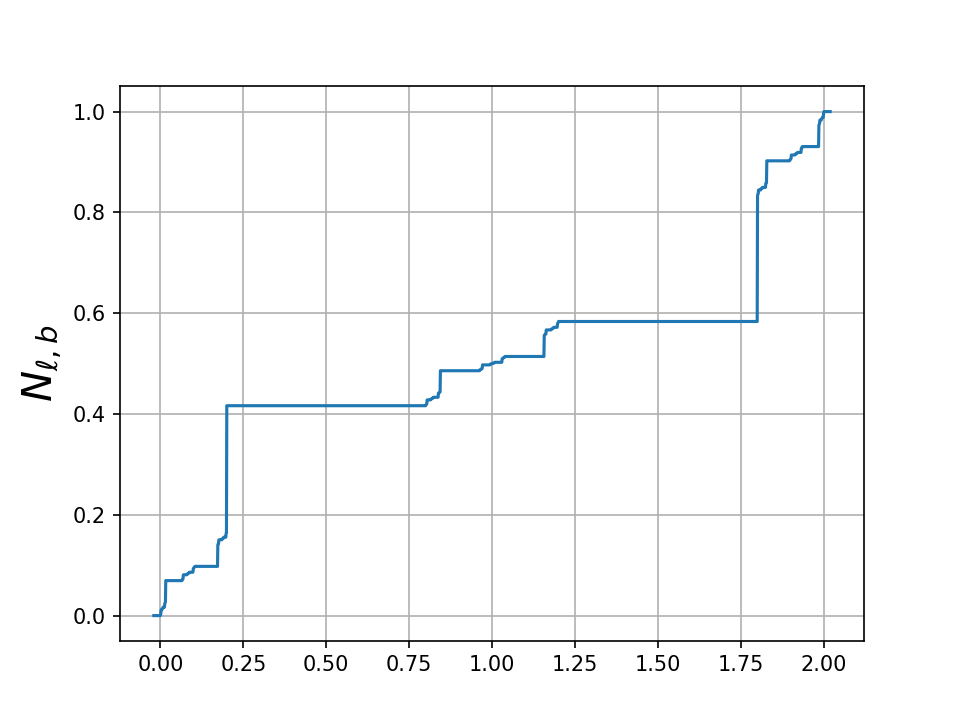} 
  \end{minipage}
  \begin{minipage}[b]{0.5\linewidth}
    \centering
    \includegraphics[width=1.0\linewidth]{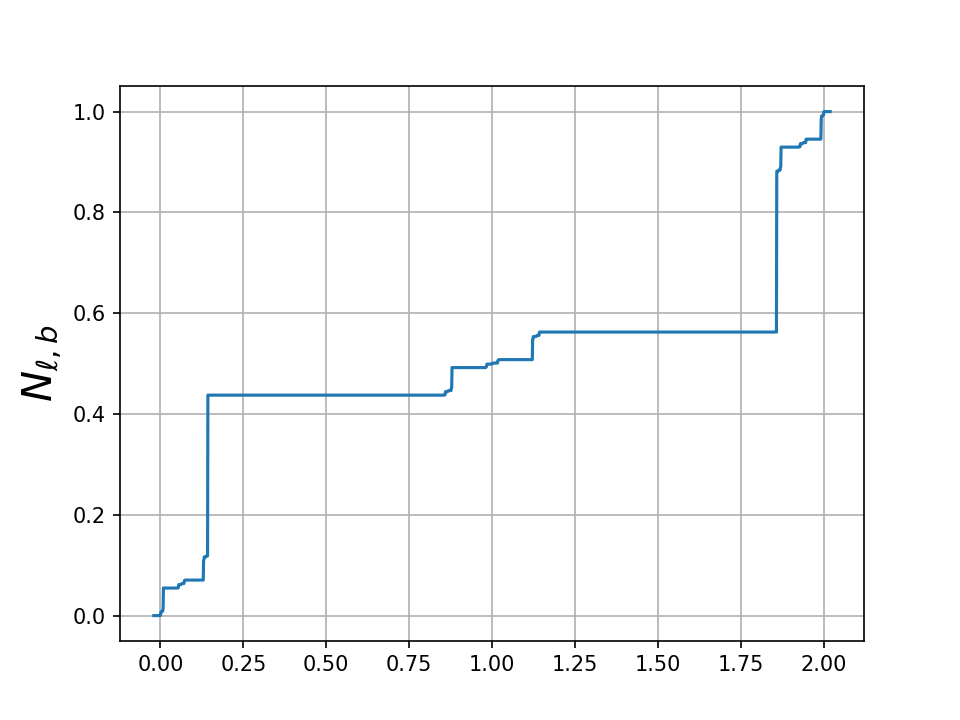} 
  \end{minipage} 
  \caption{Numerical computations of the normalized eigenvalue counting function ${N}_{\ell,b}$ for branching parameters $b=2$  (top left),  $b=3$ (top right),  $b=4$  (bottom left),  $b=6$ (bottom right). All computations are  for level $\ell=5$.}
  \label{fig:IDSdifferentParameters}
\end{figure}

We follow ideas presented in \cite[Section 5.4]{Kirsch2008RandomOperator} and define the \textit{density of states} of  $\Delta_{\infty,b}$. We start by considering the spectrum of $\Delta^D_{\ell,b}$, which consists of finitely many eigenvalues.
Recall that $\mult_{\ell}(\lambda)$ is the multiplicity of $\lambda\in\sigma(\Delta^D_{\ell,b})$. The \textit{density of states} of $\Delta^D_{\ell,b}$ is the normalized sum of Dirac measures
\begin{equation}
\label{eq:DOS}
\nu_{\ell,b}(\{\lambda\})
 =  \frac1{|V(G_{\ell})|-2} \sum_{\lambda\in\sigma(\Delta^D_{\ell,b})}\mult_\ell(\lambda) \delta_\lambda
 \end{equation}
and the \textit{normalized eigenvalue counting} function of  $\Delta^D_{\ell,b}$ is $N_{\ell,b}(x):=\nu_{\ell,b}\big((-\infty,x]\big)$.

Figure  \ref{fig:IDSdifferentParameters} depicts $N_{\ell,b}$ for different branching parameters. Proposition~\ref{prop:SpectralDiriFinite}(1) tells us the spectrum can be written as $\sigma(\Delta^D_{\ell,b})=\cup_{m=0}^{l-1}R_b^{-m}(\mathscr{E}_b)$ and Lemma~\ref{lem:dynamofRb} says these sets are disjoint.  Inserting the multiplicities from Proposition~\ref{prop:SpectralDiriFinite}(2) and using the easily verified formula $|V(G_\ell)|=2((b+2)^l+b)/(b+1)$ we obtain
\begin{equation}
	\nu_{\ell,b} = \sum_{m=0}^{\ell-1}  \frac{(b-1)(b+2)^{\ell-m-1}+2}{2((b+2)^\ell-1)} \sum_{\lambda\in R_b^{-m}(\mathscr{E}_b)} \delta_\lambda.
	\end{equation}
It follows from general dynamical systems theory that $\nu_{\ell,b}$ converges weakly as $\ell\to\infty$, however the concrete nature of the present setting allows for a sharper result, giving both the rate of convergence and a residual measure.  Before stating it we note that $\big|  R_b^{-m} (\mathscr{E}_b)  \big| = 2 \cdot 3^{m}$ and hence the following series converges in total variation:
\begin{equation}\label{eq:defnDOS}
	\nu_b =  \sum_{m=0}^\infty  \frac{b-1}{2(b+2)^{m+1}} \sum_{\lambda\in R_b^{-m}(\mathscr{E}_b)} \delta_\lambda.
	\end{equation}

\begin{theorem}
\label{prop:weakSelfSimilarity}
The measure $\nu_b$ is atomic with support precisely the Julia set $\mathcal{J}(R_b)=\sigma(\Delta_{\infty,b})$, and satisfies the functional equation
\begin{align}
\label{eq:DOSselfsimilar}
2(b+2)\nu_b = 2\nu_b\circ R_b + (b-1)\sum_{\lambda\in\mathscr{E}_b} \delta_\lambda.
\end{align}
In particular,  $\nu_b(A) = (b+2) \ \nu_b\circ S_{b,j}(A)$  for each $j=1,2,3$, where the maps $S_{b,j}$ are the branches of the inverse of $R_b$ and $A \subset [0,2]\setminus\mathcal{E}_b$. 
\end{theorem}
\begin{proof}
The first statement is obvious and the second follows from  Theorem~\ref{thm:spectrumOfinfiniteLap} and its proof.  The third needs only to be verified for atoms, on which it is immediate from the definition~\eqref{eq:defnDOS}.
\end{proof}
\begin{theorem}
\label{thm:DOSconvergence}
The sequence $\nu_{\ell,b}$ converges geometrically to $\nu_b$ as $\ell\to\infty$ in the total variation sense. More precisely, $\frac12\Bigl( \frac{b+2}3 \Bigr)^\ell (\nu_b-\nu_{\ell,b})$ converges weakly to the harmonic measure $\mu$ on $\mathcal{J}(R_b)$. The measure $\nu_b$ is therefore the density of states for $\Delta_{\infty,b}$.
\end{theorem}
\begin{remark}
Dang, Grigorchuk and Lyubich have proved a result of a similar nature for the considerably more complicated two-dimensional dynamics arising in the computation of the spectra of certain self-similar groups, see~\cite{DangGrigorchukLyubich2021}, especially Theorem~A and Remark~1.1 therein.
\end{remark}
\begin{proof}
We have already computed $|V(G_\ell)|=\frac2{b+1}\bigl((b+2)^\ell+b\bigr)$ and $\mult_\ell(\lambda)$ was given in Proposition~\ref{prop:SpectralDiriFinite}(2). Using these we compute
\begin{equation*}
	\frac{\mult_\ell(\lambda)}{|V(G_\ell)|-2} - \frac{b-1}{2(b+2)^{m+1}}
	= \frac{b-1}{2(b+2)^{m+1}((b+2)^\ell-1)} + \frac1{(b+2)^\ell-1}
	\end{equation*}
Write $\mu_m=\frac12 3^{-m} \sum_{\lambda\in R_b^{-m}(\mathscr{E}_b)}\delta_\lambda$ for the uniform probability measure on these preimages. A celebrated result of Brolin~\cite{Brolin1965} gives the weak convergence of $\mu_m$ to the harmonic measure $\mu$.  We can then write
\begin{align}
	\Bigl( \frac{b+2}3 \Bigr)^\ell (\nu_b-\nu_{\ell,b})
	&= \frac{b-1}{b+2} \frac{1}{1-(b+2)^{-l}} 3^{-l} \sum_0^{\ell-1}  \biggl(\frac3{b+2}\biggr)^m \mu_m 
	+  \frac{2}{1-(b+2)^{-l}}  \sum_0^{\ell-1} 3^{m-\ell} \mu_m \notag\\
	&\quad +  \frac{b-1}{b+2} \sum_\ell^\infty \Bigl(\frac3 {b+2}\Bigr)^{m-\ell}\mu_m \label{eq:proofofcvgewithresidualstep1}
 	\end{align}
It is easily seen that the total variation of the first term is bounded by $3^{-l}$ and hence has limit zero in this sense as $\ell\to\infty$. The weak convergence of the other two terms is routine.  Fix $f\in C_c(\mathbb{R})$ and for $\epsilon>0$ take $k$ so that $\bigl|\int f d(\mu_m-\mu)\bigr|<\epsilon$ when $m> k$ and let $C=2\sup_m\bigl|\int fd\mu_m\bigr|<\infty$. Then write
\begin{equation*}
	\Bigl( 1+ \frac{1-3^{-l}}{1-(b+2)^{-l}}\Bigr)\mu = 
	\frac{2}{1-(b+2)^{-l}}  \sum_0^{\ell-1} 3^{m-\ell} \mu + \frac{b-1}{b+2} \sum_\ell^\infty \Bigl(\frac3 {b+2}\Bigr)^{m-\ell}\mu
	\end{equation*}
and combine it with~\eqref{eq:proofofcvgewithresidualstep1}, splitting the first sum into $\sum_0^k$ and $\sum_{k+1}^{\ell-1}$, then estimating $\bigl|\int fd(\mu_m-\mu)\bigr|$ by $C$ for $0\leq m\leq k$  and by $\epsilon$ for $m>k$ to obtain:
\begin{align*}
	\lefteqn{\Biggl|\int f d\biggl( \Bigl( \frac{b+2}3 \Bigr)^\ell (\nu_b-\nu_{\ell,b}) - \Bigl( 1+ \frac{1-3^{-l}}{1-(b+2)^{-l}}\Bigr)\mu \biggr)\Biggr|}\quad&\\
	&\leq \frac{2}{1-(b+2)^{-l}}  \sum_0^{\ell-1} 3^{m-\ell} \biggl| \int fd(\mu_m-\mu)\biggr| + \frac{b-1}{b+2} \sum_\ell^\infty \Bigl(\frac3 {b+2}\Bigr)^{m-\ell} \biggl|\int fd(\mu_m-\mu)\biggr| \\
	&\leq  \frac{3^{k+1-\ell}}{1-(b+2)^{-l}} 2\sup_m\biggl|\int fd\mu_m\biggr| + \frac{2}{1-(b+2)^{-l}}  \sum_{k+1}^{\ell-1} 3^{m-\ell}\epsilon +  \frac{b-1}{b+2} \sum_\ell^\infty \Bigl(\frac3 {b+2}\Bigr)^{m-\ell}\epsilon \\
	&=  \frac{3^{k+1-\ell}}{1-(b+2)^{-l}} 2\sup_m\biggl|\int fd\mu_m\biggr| +2\epsilon
	\end{align*}
This converges to zero as $\ell\to\infty$, establishing the asserted weak convergence. The  remaining statements are immediate.
\end{proof}
\begin{corollary}
\label{coro:DefIDS}
The normalized eigenvalue counting functions $N_{\ell,b}$ converge uniformly at a geometric rate to $N_b(x):=\nu_b((-\infty,x])$.
\end{corollary}

{
As a consequence of the preceding we can compute a functional equation for the integrated density of states that implies there is a renormalized limit of the asymptotic behavior at zero.  We begin with a lemma.

\begin{lemma}\label{lem:countingfunctionforfractal}
For $n\geq1$ and $x\in[0,2]$ we have $R_b^{\circ n}\Bigl(\bigl[0,(R_b'(0))^{-n}x\bigr]\Bigr)\subset \bigl([0,x)\bigr)\cap\bigl[0,\frac1{b+1}\bigr)$.
\end{lemma}
\begin{proof}
Direct computation gives
\begin{equation*}
R_b'(z) = \frac{3(b+1)^2}{b} (z^2-2z) + \frac{(2b+1)(b+2)}b
\end{equation*}
and we find $R_b'(z)< R_b'(0)$ when $z\in(0,2)$.  The mean value theorem and $R_b(0)=0$ then provide $R_b(z)< R_b'(0)z$ for $z\in(0,2]$. We can iterate this to obtain $R_b^{\circ m}(w)< (R_b'(0))^m w$ for $1\leq m\leq n+1$ and $w\in[0,z]$ provided $(R_b'(0))^m z\in(0, 2]$ for $1\leq m\leq n$.

For $x=0$ the statement of the lemma is trivial. If $x\in(0,2]$ let $z=(R_b'(0))^{-n}x<2$. Since $R_b'(0)>1$ we have $(R_b'(0))^m z\in(0, 2]$ for $1\leq m\leq n$ and thus $R_b^{\circ m}(w)< (R_b'(0))^m w\leq 2$ for $1\leq m\leq n+1$ and $w\in[0,z]$.  Then the fact that $R_b:[0,\frac1{b+1}]\to[0,2]$ is a strictly increasing bijection implies $R_b^{\circ n}([0,z])\subset [0, x)\cap[0,\frac1{b+1})$.
\end{proof}

\begin{figure}[h]
    \centering
    \includegraphics[width=0.6\textwidth]{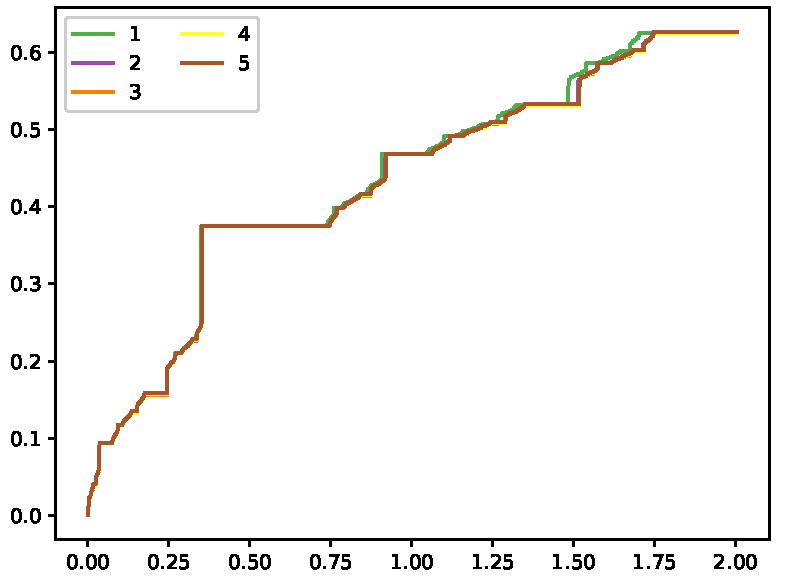}
    \caption{   Asymptotic behavior of $(b+2)^n N_b (R_b'(0))^{-n}x)$ seen by plotting for $n$ from 1 to 5, where $b=2$, and $R_b'(0)=10$. The graphs indicate proximity to a limit at low values of $n$, the reader is also referred to \cite{Sabot2000}.}
    \label{fig:sabotPic}
\end{figure}

Using the lemma we can describe the asymptotic behavior of the integrated density of states $N_b$ near zero with a functional equation.  The limiting behavior is graphed in Figure~\ref{fig:sabotPic}. In what follows we use Koenig's linearization theorem~\cite{Milnor2006,DerfelGrabnerVogl2012} to see that the renormalized composition powers $(S_{b,0}'(0))^{-l}S_{b,0}^{\circ l}$ converge on the basin of attraction of $0$ under $S_{b,0}$ to the unique holomorphic function $T_b$ that has the property, known as a Poincar\'e functional equation \cite{Grabner2015}, 
\begin{equation}\label{eqn:fnaleqnforTb}
T_b \circ S_{b,0}(z)=S_{b,0}'(0)T_b(z).
\end{equation}
We remark that this basin of attraction is the open interval from $-\infty$ to the critical point of $R_b$ in $(0,1)$, so contains $[0,2]$, and that the functional equation implies $T_b$ is strictly increasing.

\begin{theorem}\label{thm-IDS}
For $x\in[0,2]$, the function $N_b$ satisfies
 
\[ \lim_{n\to\infty} (b+2)^n N_b (R_b'(0)^{-n}x) = N_b\circ T_b^{-1}(x), \]  and hence 
\[  (b+2)N_b\circ T_b^{-1}(R_b'(0))^{-1}x) = N_b\circ T_b^{-1}(x). \]

\end{theorem}
\begin{proof}
Recall $N_b(y)=\nu_b((-\infty,y])$, so using~\eqref{eq:defnDOS} and the notation $\#A$ for the number of points in a finite set $A$ gives
\begin{equation*}\label{eqn:thmcountingforfractal1}
(b+2)^n N_b\bigl(R_b'(0)^{-n} x \bigr) 
=   \sum_{m=0}^\infty  \frac{b-1}{2(b+2)^{m-n+1}} \#\bigl( R_b^{-m}(\mathscr{E}_b) \cap [0,R_b'(0)^{-n} x]\bigr),
\end{equation*}
but Lemma~\ref{lem:countingfunctionforfractal} tells us that $R_b^{\circ n}([0,R_b'(0)^{-n} x])\subset [0,\frac1{b+1})$ which does not intersect  $\mathscr{E}_b=\{\frac1{b+1},\frac{2b+1}{b+1}\}$, so the summation terms with $0\leq m\leq n$ are zero. Moreover, $R_b^{\circ n}$ is bijective and strictly increasing on $[0,R_b'(0)^{-n} x]$, so for $m\geq n$
\begin{equation*}
\#\bigl( R_b^{-m}(\mathscr{E}_b) \cap [0,R_b'(0)^{-n} x]\bigr)
=\#\bigl( R_b^{n-m}(\mathscr{E}_b) \cap [0,R_b^{\circ n}(R_b'(0)^{-n} x)]\bigr).
\end{equation*}
Combining these observations and changing variables gives
\begin{align*}
(b+2)^n N_b\bigl(R_b'(0)^{-n} x \bigr) 
&=   \sum_{m=n}^\infty  \frac{b-1}{2(b+2)^{m-n+1}}\#\bigl( R_b^{n-m}(\mathscr{E}_b) \cap [0,R_b^{\circ n}(R_b'(0)^{-n} x)]\bigr) \notag\\
&= \sum_{k=0}^\infty  \frac{b-1}{2(b+2)^{k+1}}\#\bigl( R_b^{-k}(\mathscr{E}_b) \cap [0,R_b^{\circ n}(R_b'(0)^{-n} x)]\bigr) \notag\\
&= \nu_b \bigl( [0, R_b^{\circ n}(R_b'(0)^{-n} x)] \bigr)
= N_b\bigl( R_b^{\circ n}(R_b'(0)^{-n} x)\bigr)   \label{eqn:thmcountingforfractal2}
\end{align*}
where the penultimate step used~\eqref{eq:defnDOS}.

We wish to take the limit in~\eqref{eqn:thmcountingforfractal1}, but $N_b$ is only right continuous. However, from the proof of Lemma~\ref{lem:countingfunctionforfractal} we have that for $x\in[0,2]$
\begin{equation*}
	R_b^{\circ (n+1)}\bigl(R_b'(0)^{-(n+1)}x\bigr)
	=R_b^{\circ n} \bigl( R_b(y)\bigr)
	\leq R_b^{\circ n}\bigl( R_b'(0)y \bigr)
	= R_b^{\circ n}\bigl(R_b'(0)^{-n}x\bigr)
	\end{equation*}
so that $R_b^{\circ n}\bigl(R_b'(0)^{-n}x\bigr)$ is decreasing in $n$ and right continuity suffices for
\begin{equation}\label{eqn:thmcountingforfractal3}
\lim_{n\to\infty} (b+2)^n N_b\bigl(R_b'(0)^{-n} x \bigr) 
	=\lim_{n\to\infty} N_b \bigl(  R_b^{\circ n}(R_b'(0)^{-n} x)\bigr) 
	= N_b \bigl(\lim_{n\to\infty}  R_b^{\circ n}(R_b'(0)^{-n} x)\bigr).
\end{equation}

Finally, we identify the map $\lim_{n\to\infty}  R_b^{\circ n}(R_b'(0)^{-n}$. 
Recall from after Definition~\ref{def:gapsforcompactfractalcase} that the holomorphic map $T_b$ is the limit of $(S_{b,0}')^{-l}S_{b,0}^{l}$ on a neighborhood of $[0,2]$ and is a strictly increasing, hence invertible, function uniquely determined by the functional equation $T_ b\circ S_{b,0}(z)=S_{b,0}'T_b(z)$. The fact that $S_{b,0}$ is inverse to $R_b$ implies $ R_b^{\circ n}(R_b'(0)^{-n}x)\to T_b^{-1}(x)$, where $T^{-1}_b$ is strictly increasing and characterized by the property $R_b\circ T_b^{-1}(x)=T_b^{-1}(R_b'(0)x)$. Substitution into~\eqref{eqn:thmcountingforfractal3} completes the proof.
\end{proof}

} 

\section{Spectral Gaps and Gap Labeling Theorem}
\label{sec-GapLabeling}

A gap in the spectrum of the Laplacian $\Delta_{\infty,b}$ is simply a maximal bounded open interval that does not intersect  $\sigma(\Delta_{\infty,b})$. Not all Laplacians have gaps in their spectrum, but it is a common feature of certain self-similar graphs and fractals that has interesting consequences~\cite{StrichartzGaps,Akkermans2009ComplexDim}.  In particular, on fractals and graphs that admit spectral decimation one can detect the presence of spectral gaps using the decimation function~\cite{MR2587582,HareTeplyaev2012}; these methods are applicable to $R_b$ and they show that $\sigma(\Delta_{\infty,b})$ has gaps.  Our goal in this section is to give a more refined description of the spectral gaps  in $\sigma(\Delta_{\infty,b})$ by labeling each gap with the (constant) value attained by the normalized eigenvalue counting function $N_b$ on the gap.

\begin{figure}[htb]
\centering
\includegraphics[width=0.9\textwidth]{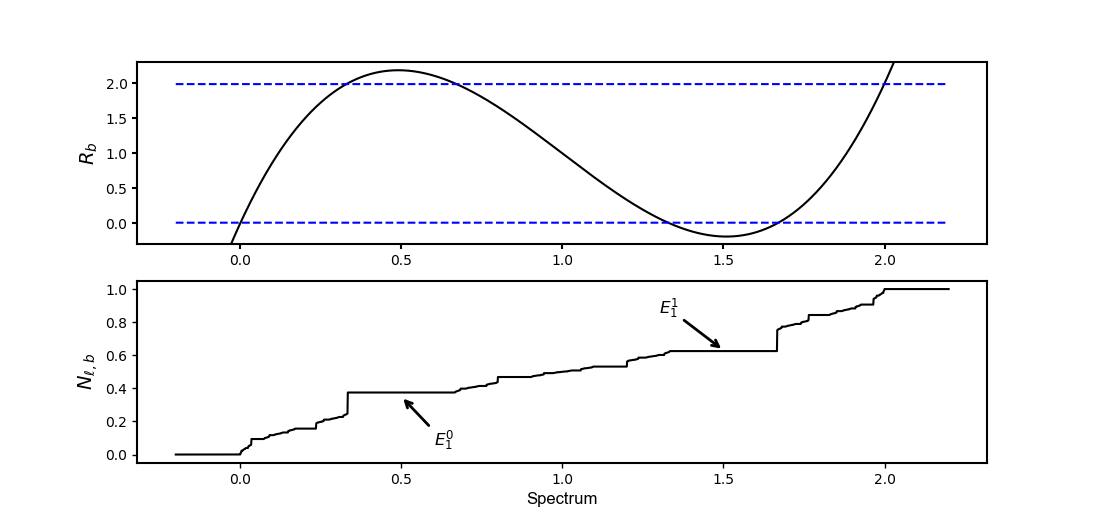}
\vspace{0cm}

\includegraphics[width=0.9\textwidth]{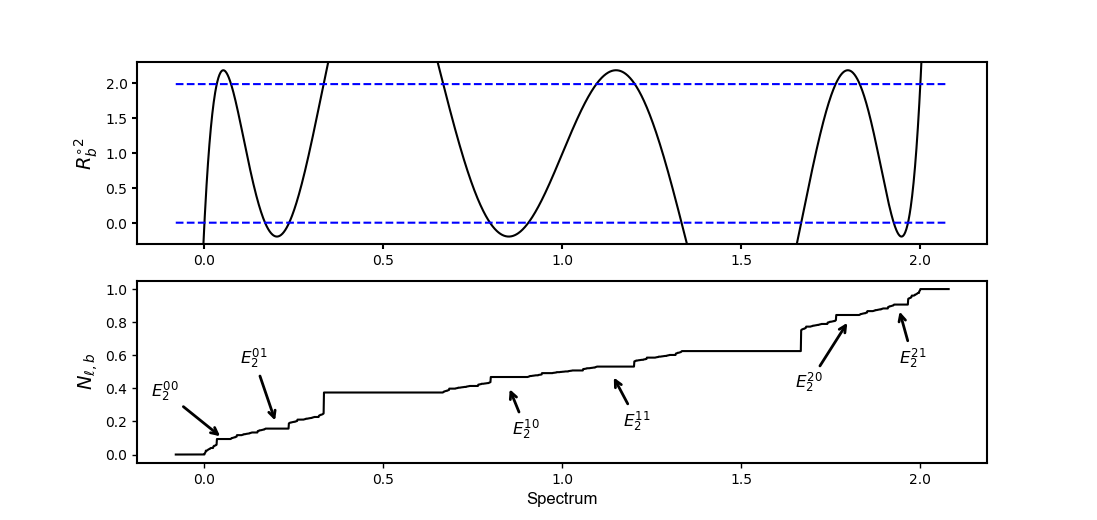}
\vspace{0cm}
\caption[The gaps labeling in the case $b=2$.]{
(Top) The spectral decimation function $R_{b}$, $b=2$, with dashed lines representing the cutoffs at $y=0$ and $y=2$, and the normalized eigenvalue counting function $N_{\ell,2}$, $\ell=6$. Note that $R_b(z)>2$  for $z \in E_1^0$ and $R_b(z)<0$ for $z \in E_1^1$, where $E_1^0$ and $E_1^1$ are the spectral gaps given in (\ref{eq:firstSpcGaps1}) and (\ref{eq:firstSpcGaps2}). 
\newline
(Bottom) The function $R^{\circ 2}_{b }(z) := R_{b }(R_{b }(z) ) $, with the dashed  lines  at $y=0$ and $y=2$, and the normalized eigenvalue counting function $N_{\ell,b}$,    $\ell=6$. Note that either $R_b(z)>2$ or $R_b(z)<0$ for $z \in E^{ij}_2$, $i,j \in \{0,1,2\}$, where the spectral gaps  $E^{ij}_2$ are defined in~\eqref{eq:spectGapsInductive}.
}
\label{fig:GeneratinggapsAMOfirstTwoGaps}
\end{figure}

The starting point of our analysis is a more detailed description of  the Julia set $\mathcal{J}(R_b)$, for which purpose we define
\begin{align}
	E_1^0&=\Bigl(\frac1{b+1},\frac b{b+1}\Bigr),  \label{eq:firstSpcGaps1} \\
	E_1^1&=\Bigl( \frac{b+2}{b+1},\frac{2b+1}{b+1}\Bigr).  \label{eq:firstSpcGaps2}
	\end{align}
These are shown in Figure~\ref{fig:GeneratinggapsAMOfirstTwoGaps} in the case $b=2$. Also recall that in  Theorem~\ref{prop:weakSelfSimilarity} we labeled the branches of the inverse of $R_b$ by $S_{b,j}$, $j=0,1,2$. It was established in Lemma~\ref{lem:dynamofRb} that the critical points of $R_b$ are in $(0,1)$ and $(1,2)$, so we can label $S_{b,j}$ to be the branch for which the domain contains $j$.  For finite words $w=w_1\dotsc w_n$ with each $w_j\in\{0,1,2\}$, further define
\begin{equation*}
	S_{b,w}=S_{b,w_1}\circ \dotsm \circ S_{b,w_n}.
	\end{equation*}

\begin{proposition}\label{prop:gapsandmaps}
The spectrum $\sigma(\Delta_{\infty,b})=\mathcal{J}(R_b)\subset[0,2]$ and has gaps exactly at the intervals $S_{b,w}(E_1^0)$ and $S_{b,w}(E_1^1)$ for $w\in \{0,1,2\}^n$, $n\in\{0\}\cup\mathbb{N}$. In particular it is totally disconnected and has Lebesgue measure zero.
\end{proposition}
\begin{proof}
From Theorem~\ref{thm:spectrumOfinfiniteLap},  $\sigma(\Delta_{\infty,b})=\mathcal{J}(R_b)$.  From the reasoning in  Lemma~\ref{lem:dynamofRb} the intervals $(-\infty,0)$ and $(2,\infty)$ are in the Fatou set of $R_b$, with the former contracting to $-\infty$ and the latter to $\infty$.   Hence $\mathcal{J}(R_b)\subset[0,2]$.   Lemma~\ref{lem:dynamofRb} also establishes that $E_1^0$ is a component of $R_b^{-1}(2,\infty)$ and $E_1^1$ is a component of $R_b^{-1}(-\infty,0)$, so are in the Fatou set. Since the endpoints of each are mapped to the repelling fixed points $2$ and $0$ (respectively) they are in $\mathcal{J}(R_b)$ so we see $E_1^0$ and $E_1^1$ are maximal bounded open intervals not intersecting $\sigma(\Delta_{\infty,b})=\mathcal{J}(R_b)$ and are therefore spectral gaps.

Again from Lemma~\ref{lem:dynamofRb} we recall that there is a local maximum of $R_b$ in $E_1^0$ and a local minimum in $E_1^1$, from which the critical points lie in the gaps and thus the restrictions of the $S_{b,j}$ to $[0,2]$ satisfy
\begin{align*}
	S_{b,0}&: [0,2]\to \Bigl[ 0,\frac{1}{b+1} \Bigr]\\
	S_{b,1}&: [0,2]\to \Bigl[ \frac{b}{b+1},\frac{b+2}{b+1} \Bigr]\\
	S_{b,2}&: [0,2]\to \Bigl[ \frac{2b+1}{b+1},2 \Bigr]
	\end{align*}
are bijective, with $S_{b,0}$ and $S_{b,2}$ being orientation preserving and $S_{b,1}$ being orientation reversing.  It follows immediately that $S_{b,w}(E_1^1)$ and  $S_{b,w}(E_1^1)$ are gaps for any choice of $w\in\{0,1,2\}^n$, $n=0,1,2,\dotsc$.

In order to show that we have found all gaps in the spectrum we need to see that the gaps we have described are the only bounded Fatou components.  Standard but somewhat sophisticated results in dynamical systems could be applied to obtain this. For instance, referring to~\cite{MR1230383}, Theorem~IV.1.3 shows any Fatou component is preperiodic, so fits the classification of Theorem~IV.2.1, but then there is a critical point either in one of the periodic components as in Chapter~III.2 or adherent to the boundary of such as in Theorem~V.1.1, neither of which is impossible because our critical points have unbounded orbits).  However an elementary argument is also available. Computing the derivative $R_b'(z)=\frac3b(b+1)^2(z-1)^2-\frac1b(b^2+b+1)$ at the points $1\pm \frac1{b+1}$ and at $1\pm \frac b{b+1}$ we find that  $|R_b'(z)|$ is bounded below by some $C>1$ on the complement of $E_1^{(0)}\cup E_1^{(1)}$. Thus any Fatou component with orbit confined to $[0,2]\setminus(E_1^{(0)}\cup E_1^{(1)})$ would strictly grow (in length) under iteration; after a finite number of iterations this would lead to the existence of a fixed point $x$ for some power $R_b^k$  in the orbit of the Fatou component, but the bound on the derivative would make $x$ a repelling fixed point, so it would be in the Julia set, leading to a contradiction.

It is now easy to see that $\sigma(\Delta_{\infty,b})=\mathcal{J}(R_b)$ is totally disconnected because inverse orbits of $0$ and $2$ under $R_b$ are both dense in the Julia set (as are inverse orbits of any point from the Julia set)  and each point in such an orbit is an endpoint of one of the gaps we have just described. This is a special case of a standard result about polynomial Julia sets for which all critical points maps into the unbounded Fatou component (Theorem~II.4.2 in~\cite{MR1230383}) and was first proved by Brolin, who also showed the more difficult fact that the Lebesgue measure of this set is zero~\cite[Theorem~13.1(2)]{Brolin1965}.
\end{proof}

It was established in Proposition~\ref{prop:gapsandmaps} that spectral gaps occur where the composition powers of $R_b$ lie outside the interval $[0,2]$. This is illustrated for $R_b$ in the top part of Figure~\ref{fig:GeneratinggapsAMOfirstTwoGaps}, which has gaps $E^0_1$ and $E^1_1$ as described in~\eqref{eq:firstSpcGaps1} and~\eqref{eq:firstSpcGaps2}. The gaps for the second composition power $R_b^{\circ 2}=R_b\circ R_b$  are shown in the bottom part of Figure~\ref{fig:GeneratinggapsAMOfirstTwoGaps}.

We define a notation for the gaps inductively, choosing it so that they are ordered from left to right in a natural fashion.  The gaps at scale $1$ are $\{E_1^0,E_1^1\}$, and supposing those of scale $k$ to have been labeled by $\{E_k^w\}$ for $w\in\{0,1,2\}^{k-1}\times\{0,1\}$ in such a manner that $E_k^w$ is to the left of $E_k^{w'}$ if and only if $\sum_1^k (w_j-w_j') 3^{-j}<0$ we let
\begin{align}
\label{eq:spectGapsInductive}
\begin{cases}
	E_{k+1}^{0w}&=S_{b,0}(E_k^w)\\
	E_{k+1}^{1w}&=S_{b,1}(E_k^{\tilde{w}})\\
	E_{k+1}^{2w}&= S_{b,2}(E_k^w)
\end{cases}
	\end{align}
where $\tilde{w}$ is the word with letters $(2-w_1)(2-w_2)\dotsm(2-w_k)(1-w_{k+1})$.
Evidently each $E_{k+1}^{jw}$ is to the left of each $E_{k+1}^{j'w}$ if $j<j'$ because the range of $S_{b,j}$ is to the left of $S_{b,j'}$.  Since $S_{b,0}$ and $S_{b,2}$ are orientation preserving, the ordering of the $E_{k+1}^{jw}$ for $w\in\{0,1,2\}^k$ is the same as that for $E_k^w$ if $j=0,2$.  For intervals $E_{k+1}^{1w}$ we note that since $S_{b,1}$ is orientation reversing, $E_{k+1}^{1w}$ is left of $E_{k+1}^{1w'}$ if and only if $E_k^{\tilde{w}}$ is right of $E_k^{\tilde{w}'}$, which occurs if and only if $\sum_{j=1}^k ((2-w_j)-(2-w'_j))3^{-j}>0$, or equivalently if $\sum_{j=1}^k (w'_j-w_j)3^{-j}<0$.  We also note that the gaps come naturally in pairs: both $E_{k+1}^{w0}$ and $E_{k+1}^{w1}$ are of the form $S_{b,w'}(E_1^0)$ and $S_{b,w'}(E_1^1)$ for some $w'$, but it is not usually the case that $w=w'$ because of the adjustment we made to ensure the left to right ordering of the gaps.

The key observation that makes it easy to obtain the gap labeling values is that  Theorem~\ref{prop:weakSelfSimilarity} provides a dynamical system on the gap labels.

\begin{theorem}\label{thm:gaplabeling}
The values taken by $N_b(x)$ on the gaps in $\sigma(\Delta_{b,\infty})$ are precisely those in the gap labeling set
\begin{equation}\label{eqn:gaplabels}
	\Biggl\{ \frac{b+1+2w_k}{2(b+2)^k}+ \frac{b+1}{2} \sum_{j=1}^{k-1}\frac{w_j}{(b+2)^j} : k\in\mathbb{N} \text{ and } (w_1,\dotsc,w_k) \in\{0,1,2\}^{k-1}\times\{0,1\} \Biggr\}.
	\end{equation}
The specific value in~\eqref{eqn:gaplabels} is attained on the scale $k$ gap $E_k^{w_1\dotsc w_k}$.
\end{theorem}

\begin{proof}
We prove inductively that the values of $N_b(x)$ on the scale $k$ gaps are given in~\eqref{eqn:gaplabels}.
From  Theorem~\ref{prop:weakSelfSimilarity} and the fact that $\nu_b$ is a probability measure we see immediately that when $k=1$
\begin{equation*}
	N_b(x)=\begin{cases}
		\frac{b+1}{2(b+2)}&\text{ if $x\in E_1^1$},\\
		\frac{b+3}{2(b+2)} &\text{ if $x\in E_1^2$}.
		\end{cases}
	\end{equation*}
Suppose that the values of $N_b(x)$ on the gaps $E_k^w$ are of the form given in~\eqref{eqn:gaplabels} and consider a scale $k+1$ gap $E_{k+1}^{lw}$.  For $x\in E_{k+1}^{lw}$ we have $x=S_{b,l}(y)$, where $y\in E_k^w$ if $l\in\{0,2\}$ and $y\in E_k^{\tilde{w}}$ if $l=1$. 

In order to compute $N_b(x)=\nu_b([0,x])$ we decompose the interval according to the range of the functions $S_{b,l}$ and write it in terms of $\nu_b\circ S_{b,l}$.  Note that the second equality involves the orientation-reversal property of $S_{b,1}$. We then apply Theorem~\ref{prop:weakSelfSimilarity}, noting that since $y$ lies in a gap it is not in $\mathcal{E}_b$, to obtain
\begin{align}
	\nu_b([0,x]) 
	&=\begin{cases}
		\nu_b\bigl([0,S_{b,0}(y)]\bigr) &\text{ if $l=0$},\\
		\nu_b\biggl( \bigl[0,\frac{b}{b+1}\bigr]\biggr) + \nu_b \biggl( \bigl[\frac{b}{b+1}, S_{b,1}(y)\bigr]\biggr) &\text{ if $l=1$},\\
		\nu_b\biggl( \bigl[0,\frac{2b+1}{b+1}\bigr]\biggr) + \nu_b \biggl( \bigl[\frac{2b+1}{b+1},S_{b,2}(y)\bigr]\biggr) &\text{ if $l=2$},
	 	\end{cases} \notag\\
	&= \frac{l(b+1)}{2(b+2)} + \begin{cases}
		 \nu_b\circ S_{b,0}( [0,y]) &\text{ if $l=0$}, \\
		 \nu_b\circ S_{b,l} \bigl( [y,2] \bigr)&\text{ if $l=1$ },\\
		 \nu_b\circ S_{b,2} \bigl([0,y] \bigr) &\text{ if $l=0$}, 
		\end{cases} \notag\\
	&=\frac{l(b+1)}{2(b+2)} + \frac1{b+2} \begin{cases}
									 \nu_b( [0,y])  &\text{ if $l\in\{0,2\}$}, \\
									 \nu_b \bigl( [y,2] \bigr)&\text{ if $l=1$.}
									\end{cases} \label{eqn:measurecomputationforgaplabels}
	\end{align}

When $l\in\{0,2\}$ the inductive hypothesis~\eqref{eqn:gaplabels} provides a formula for $\nu_b([0,y])$. For the $l=1$ case we must instead use this hypothesis to write
\begin{align*}
	 \nu_b \bigl( [y,2] \bigr)
	&=  \nu_b \bigl( [ 0, 2] \bigr) - \nu_b \bigl( [0,y) \bigr) \\
	&=1-    \Biggl(  \frac{b+1+2\tilde{w}_k}{2(b+2)^k} +   \frac{b+1}2\sum_{j=1}^{k-1} \frac{\tilde{w}_j}{(b+2)^j }\Biggr)\\
	&=1-   \Biggl(  \frac{b+3-2w_k}{2(b+2)^k} +  \frac{b+1}2 \sum_{j=1}^{k-1} \frac{2-w_j}{(b+2)^j} \Biggr)\\
	&=  \frac{b+1+2w_k}{2(b+2)^k}+ \frac{b+1}{2} \sum_{j=1}^{k-1}\frac{w_j}{(b+2)^j}
	\end{align*}
by computing the series. Note that in the second equality it was important that $y$ lies in a gap, as this ensures the endpoint $S_{b,1}(y)$ is not the location of a Dirac mass of $\nu_b$ and thus $\nu_b \bigl( [0,y) \bigr)=\nu_b \bigl( [0,y] \bigr)$.  

We now have the same formula~\eqref{eqn:gaplabels} for both $\nu_b[0,y]$ if $l\in\{0,2\}$ and $\nu_b([y,2])$ when $l=1$; substituting into~\eqref{eqn:measurecomputationforgaplabels} and observing that $w_j$ is the $(j+1)^\text{th}$ letter of the word $lw$ for the gap $E_{k+1}^{lw}$ closes the induction.
\end{proof}

One way to view~\eqref{eqn:measurecomputationforgaplabels} is that it expresses the range of the counting function $N_b(x)$ is a self-similar set in $[0,1]$ under the action of the three maps $Q_{0,b}(y)=\frac{y}{b+2}$, $Q_{1,b}(y) =\frac12+\frac{(1-2y}{2(b+2)}$ and $Q_{2,b}(y)=1+\frac{y-1}{b+2}$, all of which scale by the same factor $\frac1{b+2}$ and have fixed points $0,\frac12,1$ respectively; note that $Q_{0,b}$ and $Q_{2,b}$ are orientation-preserving and $Q_{1,b}$ is orientation reversing, but that this latter has little effect because the self-similar system is invariant under the reflection fixing $1$ and exchanging $0$ with $2$.  Then the observation that the values attained by $N(x)$ are limits of the gap values also gives a description of the range of $N(x)$.

\begin{corollary}\label{cor:dynamicalgaplabeling}
	The gap labels are the orbit of $\bigl\{\frac{b+1}{2(b+2)},\frac{b+3}{2(b+2)} \bigr\}$ under the iterated function system $\mathcal{Q}=\{Q_{0,b},Q_{1,b}, Q_{2,b}\}$ and the range of the counting function $N_b(x)$ is the invariant set of $\mathcal{Q}$.
	\end{corollary}


\section{Connection to the compact and non compact  fractal cases} \label{sec:Sabotsection0}

\subsection{Explicit spectrum computation for the compact fractal case}\label{sec:Sabotsection}
In the preceding sections we have considered the spectra of Laplacians on an unbounded sequence of graphs and their limits.  There is also a natural sense in which one can renormalize so that the sequence of graphs converges to a compact limit, along with an associated convergence of the Laplacian and its spectrum. In the present situation this defines a class of examples that fit entirely within the theory developed by Kigami~\cite{KigamiBook2001}. Our only purpose here is to consider the explicit structure of gaps in this context (see \cite{HareTeplyaev2012,StrichartzGaps}), so we give a minimum of details and refer frequently to~\cite{KigamiBook2001}.

In giving the basic theory we suppose $b$ to be fixed and suppress the subscript.
Recall the Laplacian $\Delta_l$ on the graph $G_l$ defined in~\eqref{eq:probabilisticGraphLaplacian} and the inner product $\langle f|g\rangle_l=\sum_{V(G_l)}f(p)g(p)\deg_l(p)$. It is easily checked that $E_l(f,g)=-\langle \Delta_{l,b}f|g\rangle=\sum (f(p)-f(q))(g(p)-g(q))$, with the sum being over all edges of $G_l$. This defines a  quadratic form on $G_l$. Computing the effective resistance between the vertices $V(G_0)$ in $G_1$ one finds that the forms $\bigl(\frac{2b+1}b\bigr)^l E_l(f,f)$ are a compatible sequence in the sense of~\cite[Definition~2.2.1]{KigamiBook2001} which, by virtue of~\cite[Theorem~2.3.10]{KigamiBook2001}, converge to a resistance form $E$ on the countable set $V_*=\cup_l V(G_l)$ that extends to the resistance completion $K$.  To realize $K$ as a self-similar set, observe that the construction of $G_l$ by replacing edges of $G_1$ with copies of $G_{l-1}$ defines for the $j^\text{th}$ edge of $G_1$ a map $G_{l-1}\to G_l$, each such map contracts resistance (this uses the finite ramification structure), and that (in the obvious manner) increasing $l$ defines an extension of the map. It follows that each such sequence of maps is eventually constant on any point of  $V_\ast$, with the limit therefore defining a self-map $\Theta_j:V_*\to V_*$ that extends continuously to an injection $\Theta_j:K\to K$.  Under these maps it is evident that the form $E$ is self-similar.  Moreover we can endow $K$ with its unique equally weighted self-similar probability measure $\mu$ and note that if we divide the degree weights used to define the inner product on $G_l$ by $2(b+2)^{l-1}$ so as to obtain a probability measure then this sequence of measures converges to $\mu$. The resulting measure is Radon, and applying~\cite[Theorems~2.4.1 and 2.4.2]{KigamiBook2001} we find that $E$ is a Dirichlet form on $L^2(\mu)$ and the associated self-adjoint Laplacian defined by $E(f,g)=-\langle \Delta_K f|g\rangle_\mu$ for $f,g\in\dom(E)$ has compact resolvent. This last fact ensures $\sigma(\Delta_K)$ has a discrete  spectrum, consisting of isolated   eigenvalues of finite multiplicity accumulating only at $\infty$.

Our goal  in this subsection is to show that the gap structure described in Section~\ref{sec-spectralDeci} has an analogue for $\sigma(\Delta_K)$.  Evidently the definition of a gap cannot be the same: for the fractal the spectrum is discrete, so the fact that it omits open intervals is trivial.  The ``correct'' definition, proposed by Strichartz~\cite{StrichartzGaps}, is that there must be a sequence of omitted intervals of size comparable to the nearby eigenvalues. Note that although~\cite{HareTeplyaev2012} proves the existence of gaps, here we present an explicit computation of the entire spectrum, including the gaps, on our class of examples. 

\begin{definition}\label{def:gapsforcompactfractalcase}
If $\lambda_n$ is the sequence of eigenvalues of a non-positive definite self-adjoint operator with compact resolvent, we say that the spectrum has gaps, or a sequence  of exponentially large gaps, if there is a sequence $N_m$ with the property that  the gap label $\inf_m\bigl(\frac{\lambda_{N_m+1}}{\lambda_{N_m}} -1\bigr)$ is bounded away from zero. The corresponding sequence of gaps consists of the intervals between $\lambda_{N_m}$ and $\lambda_{N_m+1}$.
\end{definition}

Now we give a description of the spectrum of $\Delta_K$ using the spectral decimation techniques of Section~\ref{sec-spectralDeci}. We know that for $z_{l-1}$ an eigenvalue of $\Delta_{l-1}$ we can extend the corresponding eigenfunction $f_{l-1}$ to $G_l$ in such a manner that we obtain an eigenfunction $f_l$ of $\Delta_l$ with eigenvalue $z_l$ chosen from $\{S_{b,j}(z): j=0,1,2\}$, where we recall these maps are the inverse branches of $R_b$.  Thus $E_l(f_l,g)=-z_l\langle f_l,g\rangle_l$ for each $l$ and any continuous $g$ on $K$.  Re-writing this to incorporate the energy and mass renormalization scalings we have
\begin{equation}\label{eqn:spectraldecimationrenormalizes}
	\Bigl(\frac{2b+1}b \Bigr)^l E_l(f_l ,g) =  2\Bigl(\frac{(2b+1)(b+2)}{b}\Bigr)^l z_l  \frac{\langle f_l,g\rangle_l}{2(b+2)^l}.
	\end{equation}

If the energy and inner product terms are to converge in~\eqref{eqn:spectraldecimationrenormalizes} we need $z_l\to0$, for which we must require that $z_l=S_{b,0}z_{l-1}$ for all but finitely many $l$. Under that hypothesis and observing that $\frac{(2b+1)(b+2)}{b}=R_b'(0)=(S_{b,0}'(0))^{-1}$ we can use \eqref{eqn:fnaleqnforTb}. 
Then,  if $z_l=S_{b,0}(z_{l-1})$ for $l\geq k$ we have
\begin{equation}\label{eqn:convergeofTbzk}
	2\Bigl(\frac{(2b+1)(b+2)}{b}\Bigr)^l z_l \to 2\Bigl(\frac{(2b+1)(b+2)}{b}\Bigr)^k T_b(z_k).
	\end{equation}
and~\eqref{eqn:fnaleqnforTb} allows us to assume that $z_k\neq S_{b,0}(z_{k-1})$ (this includes the possibility that $k=0$).

Using the same condition, that  $z_l=S_{b,0}z_{l-1}$ for all $l>k$ implies $z_l\to0$, we discover that taking the piecewise harmonic extension of $f_l$ to $K$ defines an equicontinuous sequence; since the sequence is eventually constant on the dense set $V_*$ there is a unique limit $f$ and the harmonic extensions of the $f_l$ converge uniformly to $f$. Then the same can be said about $f_lg\to fg$ and, since the renormalized measures converge weakly to $\mu$, \eqref{eqn:spectraldecimationrenormalizes} becomes
\begin{equation}\label{eqn:formulaforevalofDelta}
	E(f,g) = \lim_l \Bigl(\frac{2b+1}b \Bigr)^l E_l(f_l ,g) = 2\Bigl(\frac{(2b+1)(b+2)}{b}\Bigr)^k T_b(z_k) \langle f,g\rangle_\mu
	\end{equation}
so that $f$ is an eigenfunction of $\Delta_K$.  By careful examination of the localized eigenfunction construction from Theorem~\ref{thm:purepointSpectrum} one checks that the eigenfunctions constructed by this method are dense in $L^2(\mu)$, so this approach finds the whole spectrum of $\Delta_K$.

\begin{theorem}\label{thm:spectrumforDeltaonK}
The spectrum of $\Delta_K$ is
\begin{equation*}
	\sigma(\Delta_K) = \bigl\{ 0, 2T_b(2) \bigr\} \cup \bigcup_{k=1}^\infty \Biggl\{  2 \Bigl( \frac{(2b+1)(b+2)}{b}\Bigr)^k T_b(z): z\in \sigma(\Delta_{k,b})\cap\bigl[ \frac b{b+1},2\bigr] \Biggr\}
	\end{equation*}
with the multiplicity of each eigenvalue given by that of $z\in\sigma(\Delta_{k,b})$ as in Proposition~\ref{prop:SpectralDiriFinite}. The spectrum has large gaps in the sense of Definition~\ref{def:gapsforcompactfractalcase}. 
To every gap $(g_1,g_2)$ in $\sigma(\Delta_{\infty,b})\cap \bigl[ \frac 1{b+1},2\bigr]$ there is a sequence of eigenvalues of $\Delta_K$ defining gaps in the sense of  Definition~\ref{def:gapsforcompactfractalcase} and its gap label is $\frac{T_b(g_2)}{T_b(g_1)}-1>0$.
\end{theorem}
\begin{proof}
The eigenvalues we constructed had values as in~\eqref{eqn:formulaforevalofDelta}, under the hypothesis that $z_k\neq S_{b,0}(z_{k-1})$ but $z_l=S_{b,0}z_{l-1}$ for all $l>k$. Then either $k=0$ and $z_0\in\{0,2\}$ or $k\geq1$ and $z_k\in\sigma(\Delta_{k,b})\cap\bigl[ \frac b{b+1},2\bigr]$.  This gives the formula for the spectrum.

Now from~\eqref{eqn:fnaleqnforTb} we see that 
\begin{equation*}
	2 \Bigl(  \frac{(2b+1)(b+2)}b\Bigr)^k T_b(2) = \Bigl( \frac{(2b+1)(b+2)}b\Bigr)^{k+1} T_b \Bigl(\frac1{b+1}\Bigr) 
	\end{equation*}
and since $T_b$ is strictly increasing this shows that the intervals in which the spectrum lies are separated by gaps of the form
\begin{equation*}
	\Biggl(  2 \Bigl( \frac{(2b+1)(b+2)}b\Bigr)^k T_b \Bigl(\frac1{b+1}\Bigr),  2 \Bigl( \frac{(2b+1)(b+2)}b\Bigr)^k T_b \Bigl(\frac b{b+1}\Bigr) \Biggr)
	\end{equation*}
Moreover, these provide gaps in the sense of Definition~\ref{def:gapsforcompactfractalcase} because taking the largest eigenvalue to the left of each of these intervals to be $\lambda_{N_m}$ one has $\inf_m\frac{\lambda_{N_m+1}}{\lambda_{N_m}}-1 = \frac{T_b(\frac b{b+1})}{T_b(\frac1{b+1})}-1$.

At the same time we see that each of the other gaps $(g_1,g_2)$ in the spectrum $\sigma(\Delta_{\infty,b})\cap \bigl[ \frac b{b+1},2\bigr]$ is replicated in the spectrum of $\Delta_K$ as a gap in the sense of  Definition~\ref{def:gapsforcompactfractalcase}. Taking $z_{N_m}$ to be the largest eigenvalue of $\Delta_{m,b}$ less than $g_1$ and $\lambda_{N_m}=2 \bigl(  \frac{(2b+1)(b+2)}b\bigr)^m T_b(z_{N_m})$, we compute from the description of $\sigma(\Delta_K)$ and the fact that $(g_1,g_2)$ is a maximal open interval in the complement of $\sigma(\Delta_{\infty,b})$, that $\inf \frac{\lambda_{N_m+1}}{\lambda_{N_m}}-1=\frac{T_b(g_2)}{T_b(g_1)}-1$, which is positive because $T_b$ is strictly increasing.
\end{proof}

\subsection{Spectrum for non-compact fractal blow-ups} \label{sec:Sabotsection2}
{

In this subsection we describe the spectrum and  the integrated density of states for an infinite blowup of the compact fractal from Section~\ref{sec:Sabotsection}, primarily using results from~\cite{StrichartzTeplyaev2012}. The blowup is made in the sense of~\cite{StrichartzFractalsinthelarge,Barlow1988,StrichartzTeplyaev2012,Affine}. This is a variant of existence results from~\cite{Sabot2000,PP1,PP2} and references therein.  For convenience we fix $b$ throughout, and only refer to it in discussing the maps $R_b$ and sets $\mathscr{E}_b$, though it affects all aspects of the construction.
In Section~\ref{sec:Sabotsection} we defined a compact fractal limit $K$ of the rescaled bubble diamond graphs that could be described as a self-similar set $K=\cup_{k=1}^{b+2}\Theta_k K$ and is endowed with a self-similar measure $\mu$ and Dirichlet form $E$ on $L^2(\mu)$.  In~\cite{StrichartzFractalsinthelarge} Strichartz defined a fractal blowup of such a $K$ for each choice of infinite sequence $\{k_\ell\}_0^\infty$  with each $k_\ell \in\{1,\dotsc,b+2\}$ by $\hat{K}=\cup_{\ell=1}^\infty \Theta_{k_1}^{-1} \circ \Theta_{k_2}^{-1}\circ \dotsm \circ \Theta_{k_\ell}^{-1}K$.  Evidently $\hat{K}$ can be thought of as a countable union of copies of $K$, and we can extend $\mu$  to $\hat{\mu}$ on $\hat{K}$ by requiring its restriction to each copy to be a copy of $\mu$. We can extend $E$ to $\hat{E}$ in a similar same manner; the domain consists of continuous functions in $L^2(\hat{K},\hat{\mu})$ with the property that the restriction to each copy of $K$ is in the domain of $E$ and that the sum of the $E$ energies over the copies of $K$ is finite.  One finds that then $\hat{E}$ is a Dirichlet form on $L^2(\hat{\mu})$ \cite{Affine}.

Comparing the preceding with Definition~\ref{def:infiniteBubbleDiamond} it is apparent that the infinite bubble diamond graph $G_\infty$ corresponding to the sequence $\{k_\ell\}$ is the blowup of the  graph $G_0$, and that one may similarly define the blowups of each graph $G_j$, which we will denote $\hat{G}_j$. Each $\hat{G}_j$ is an infinite graph and we see as we did in Section~\ref{sec:Sabotsection} that if $\hat{E}_j$ is the quadratic form obtained from the graph Laplacian on the blowup $\hat{G}_j$ then  $\bigl(\frac{2b+1}b\bigr)^j\hat{E}_j$ forms a compatible sequence of resistance forms that converges to $\hat{E}$.  Similarly, the degree weights on the blowup $\hat{G}_j$ can be divided by $2(b+2)^{j-1}$ to give a sequence of measures $\hat{\mu}_j$ that converges to $\hat{\mu}$, simply because on each copy of $K$ we are doing the computation from  Section~\ref{sec:Sabotsection}.

From the preceding we see it is possible to obtain $\hat{K}$ and the structures $\hat{\mu}$ and $\hat{E}$ either as a countable union of copies of $(K,\mu,E)$ (with appropriate requirements for the domain of $E$) or as a limit of refinements of the infinite bubble graphs with measure and energy obtained as limits of appropriately renormalized vertex measures and graph energies.  In this sense the blowup $\hat{K}$ incorporates both notions of the limit of a sequence of bubble graphs.  We let $\hat{\Delta}$ denote the Laplacian corresponding to $\hat{E}$ on $L^2(\hat{\mu})$.

Let us now consider the problem of describing the spectrum of the Laplacian corresponding to the Dirichlet form $\hat{E}$ on $L^2(\hat{\mu})$ using this analysis.   To do so, we first use the interpretation that the blowup is a countable union of copies of $K$ glued at vertices according to the graph structure of $G_\infty$ (which incorporates the dependence on the blowup sequence $\{k_\ell\}$), and that this gluing is locally finite because the number of copies of $K$ that are glued at a vertex is at most $b+1$.   This construction falls within the analysis made   in~\cite{StrichartzTeplyaev2012}, similarly to \cite{Affine}, and their work shows that one can spectrally decimate between a finite stage of the blowup, with a finite number of copies of $K$ glued at vertices of $G_n$, and the same construction for level $G_{n+1}$, using the same method as was applied in Section~\ref{sec-spectralDeci}. Taking the limit over $n$ one finds that the Laplacian on the blowup admits a spectral decimation operation exactly akin to that obtained on the graph limit $G_\infty$, except that the role of the exceptional set $\mathscr{E}_b$ is played by the spectrum of the Laplacian on $K$, and consequently we have,  (compare to the proof of Theorem~\ref{thm:spectrumOfinfiniteLap})
\begin{equation}
\sigma(\Delta_{\hat{K}}) =     \bigcup_{k=0}^\infty \left\{    \Bigl( \frac{(2b+1)(b+2)}{b}\Bigr)^{-k}   \sigma(\Delta_{K})  \right\}=     \bigcup_{k=-\infty}^\infty \left\{    \Bigl( \frac{(2b+1)(b+2)}{b}\Bigr)^{k}   \sigma(\Delta_{K})  \right\}.  
\end{equation}
Alternatively, we could have found this by considering the blowup to be a limit of the infinite graphs $\hat{G}_j$ obtained by blowing up each $G_j$ via the sequence $\{k_\ell\}$.  Specifically, if we let $\hat{\Delta}_j$ denote the Laplacian for $(\hat{G}_j,\hat{E}_j,\hat{\mu}_j)$ then the spectral decimation operation discussed in the previous subsection is applicable; an eigenvalue $\hat{z}_{j-1}$ for $\hat\Delta_{j-1}$ gives three eigenvalues for $\hat\Delta_j$, namely $S_{b,m}(\hat{z}_{j-1})$, $m=0,1,2$, and these satisfy~\eqref{eqn:spectraldecimationrenormalizes}.  Continuing the same line of reasoning via~\eqref{eqn:fnaleqnforTb} and~\eqref{eqn:convergeofTbzk} one finds that the spectrum of $\hat\Delta$ is as in Theorem~\ref{thm:spectrumforDeltaonK} except that $z_k$ may range over the spectrum of the Laplacian $\hat\Delta_k$ of the blowup $\hat{G}_k$ rather than just $\sigma(\Delta_{k,b})$ for the finite graph $G_k$. We can then insert the description of the blowup spectrum from  Theorem~\ref{thm:spectrumOfinfiniteLap}) as follows:
\begin{equation}
\sigma(\Delta_{\hat{K}}) =      T^{-1}_b(\Delta_{\infty,b}) .
\end{equation}
In both cases the integrated density of states is defined as a limit of measures computed from finite graphs and is given in Theorem~\ref{thm-IDS}.

}  

\subsection{Connection to the work of Rammal}

\begin{figure}[h]
	\begin{minipage}[b]{0.5\linewidth}
		\centering
		\includegraphics[width=0.52\linewidth]{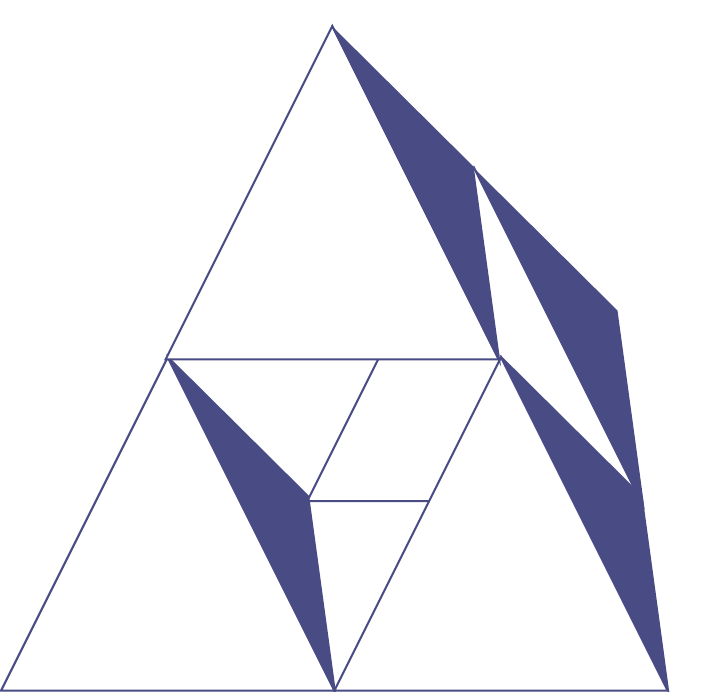} 
	\end{minipage}
	\begin{minipage}[b]{0.5\linewidth}
		\centering
		\includegraphics[width=0.52\linewidth]{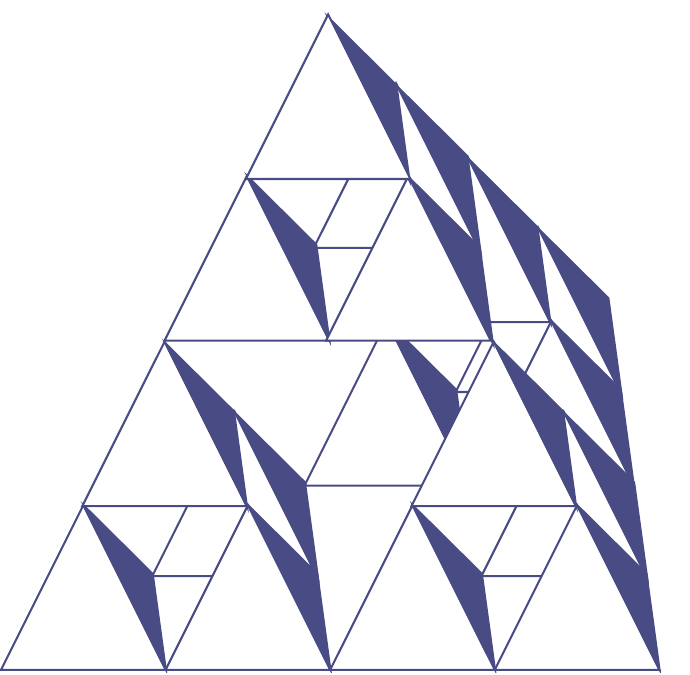} 
	\end{minipage} 
	\caption{Construction of higher dimensional Sierpinski lattices that appeared in classical papers  \cite{Rammal1983,RammalToulouse1983,Rammal1984,ShimaFukushima1992,Kigami1989,Kigami1994}.}
	\label{fig:hSG}
\end{figure}

An early motivation for our research was connecting the classical work of Rammal~\cite{Rammal1984} to the theory of gap labeling. Rammal analyzed the points of discontinuity of the normalized eigenvalue counting function for a higher dimensional Sierpinski lattice, see Figure \ref{fig:hSG},  and computed its values on the first few spectral gaps. The connection to the present work  may be seen by noting that an analogue of Theorem~\ref{prop:weakSelfSimilarity} is valid for Sierpinski lattices because these graphs admit spectral decimation, and therefore the computations of Rammal can be reproduced in the following manner.

\begin{figure}[htb]
	\centering
	\includegraphics[width=0.7\linewidth]{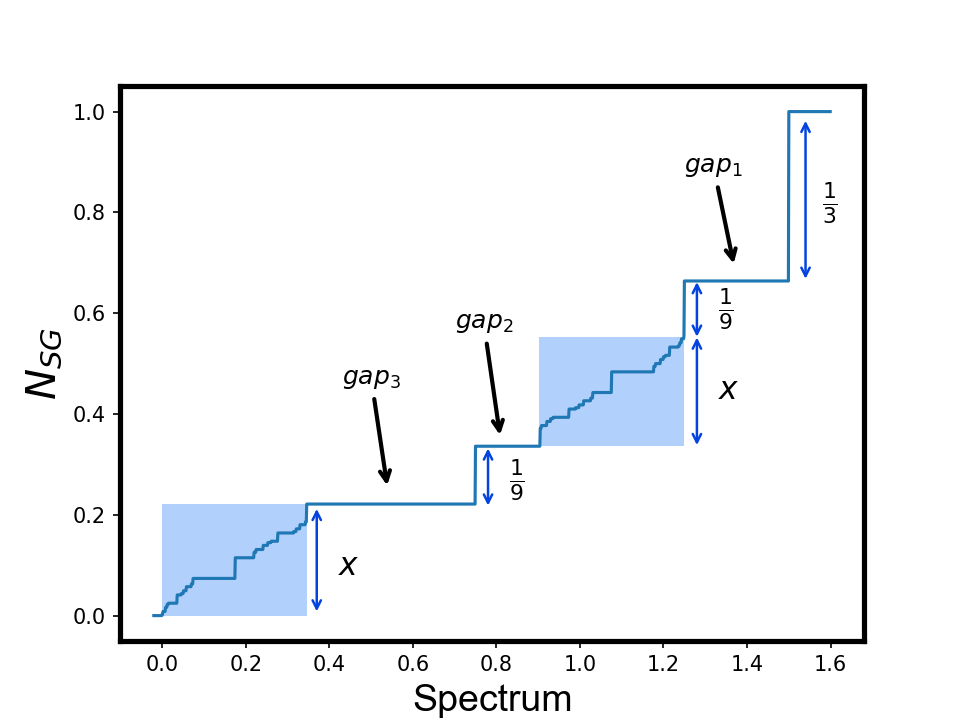} 
	\caption{ {Numerical computation of the normalized eigenvalue counting function ${N}_{SG}$ for a Sierpinski lattice.}
	}
	\label{fig:IDSforSGRammaln}
\end{figure}

\begin{figure}[h]
    \centering
    \includegraphics[width=0.7\textwidth]{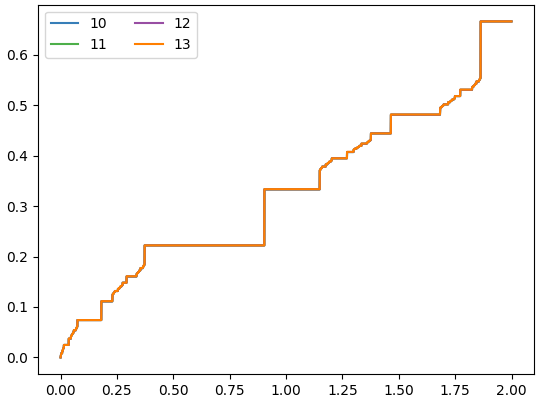}
    \caption{{
    Asymptotic behavior of $c_1^mN(c_2^mx)$ for the Sierpinski Graph at level 22, where $m$ ranges from 10 to 13 and $c_1= 3$, $c_2=\frac{1}{5}$. Since this is a high level of the graph with a high $m$, it gives a good picture of the convergence. The reader is also referred to \cite{Sabot2000}.
    }}
    \label{fig:my_labelz}
\end{figure}

The spectral decimation function associated with a probabilistic graph Laplacian $\Delta_{SG}$ on a Sierpinski lattice is a polynomial of degree two and given by $R_{SG}(z) = z(5-4z)$, see for instance~\cite{BajorinVibrationSpectra2008}. Let $S_0$ and $S_1$ denote the branches of the inverse $R_{SG}^{-1}$. The spectrum $\sigma(\Delta_{SG})$ is a Cantor set, contained in $\big[0, \frac{3}{2}\big]$ and the first two rescaled copies are (roughly speaking) given by $S_0(\sigma(\Delta_{SG}))$ and $S_1(\sigma(\Delta_{SG}))$, see the colored boxes in Figure~\ref{fig:IDSforSGRammaln}. Now the key idea of Theorem \ref{prop:weakSelfSimilarity} is that the density of states measure $\nu_{SG}$ have equal weights on these two rescaled copies; more precisely, $\nu_{SG}\big(  S_0 \big(\sigma(\Delta_{SG})\setminus \{0, {3}/{2}\} \big)  \big) =\nu_{SG}\big(  S_1 \big(\sigma(\Delta_{SG})\setminus \{0, {3}/{2}\}  \big)  \big)=x$. On the other hand, the jump discontinuities of the normalized eigenvalue counting function $N_{SG}$ were computed in \cite[Corollary 5.2]{BajorinVibrationSpectra2008} to have sizes $1/9$, $1/9$ and $1/3$ as shown in Figure~\ref{fig:IDSforSGRammaln}. Using these values it is apparent that
\begin{align*}
 \  N_{SG}(\text{gap}_3)=x=\frac{2}{9},   \quad \quad  N_{SG}(\text{gap}_2)=\frac{3}{9}, \quad \quad  N_{SG}(\text{gap}_1)=\frac{2}{3}.
\end{align*}
This method may be iterated to compute $N_{SG}$ on the spectral gaps occurring at successive scales, as illustrated in Figure~\ref{fig:my_labelz}. 

Additional  examples to which this approach may be applied may be found in~\cite{BajorinVibration3Ngasket2008,BajorinVibrationSpectra2008} or, in the context of almost Mathieu operators,~\cite{BaluMogOkoTep2021spectralAMO}.

\subsection*{Acknowledgments} 
This research was supported in part by 
the University of Connecticut Research Excellence Program, 
by DOE grant DE-SC0010339 and by NSF DMS grants 
1613025, 
1659643, 
1950543, 
2008844. 
The work of G.~Mograby was additionally supported by ARO grant W911NF1910366.

\bibliographystyle{plain}
\bibliography{BibListGapLabeling2022}


\end{document}